\documentclass[conference]{IEEEtran}
\IEEEoverridecommandlockouts
\usepackage{cite}
\usepackage{amsmath,amssymb,amsfonts}
\usepackage{algorithmic}
\usepackage{graphicx}
\usepackage{textcomp}
\usepackage{xcolor}
\usepackage{tabularx}

\usepackage{caption}
\usepackage{subcaption}
\usepackage{float}
\usepackage{booktabs}
\usepackage{hhline}
\usepackage{colortbl}
\usepackage{multirow}

\usepackage{ulem}
\usepackage{authblk}

\newif\ifdraft
\drafttrue
\ifdraft
\newcommand{\todo}[1]{ {\textcolor{red} { ***TODO: #1 }}}

\else
\newcommand{\todo}[1]{}
\fi

\def\BibTeX{{\rm B\kern-.05em{\sc i\kern-.025em b}\kern-.08em
    T\kern-.1667em\lower.7ex\hbox{E}\kern-.125emX}}
\begin{document}

\title{Leveraging User Access Patterns and Advanced Cyberinfrastructure to Accelerate Data Delivery from Shared-use Scientific Observatories}

\author[1]{Yubo Qin}
\author[1]{Ivan Rodero}
\author[1]{Anthony Simonet}
\author[2]{Charles Meertens}
\author[2]{Daniel Reiner}
\author[2]{James Riley} 
\author[1,3]{Manish Parashar}
\affil[1]{Rutgers Discovery Informatics Institute, Piscataway, New Jersey, USA}
\affil[2]{UNAVCO, Boulder, Colorado, USA}
\affil[3]{Scientific Computing Imaging Institute, University of Utah, Salt Lake City, Utah, USA}
\affil[ ]{\{yubo.qin; irodero; anthony.simonet; parashar\}@rutgers.edu}
\affil[ ]{\{chuckm; reiner; jriley\}@unavco.org}



\maketitle

\begin{abstract}

With the growing number and increasing availability of shared-use instruments and observatories, observational data is  becoming an essential part of application workflows and contributor to  scientific discoveries in a range of disciplines. However, the corresponding growth in the number of users accessing these facilities coupled with the expansion in the scale and variety of the data, is making it challenging for these facilities to ensure their data can be accessed, integrated, and analyzed in a timely manner, and is resulting significant demands on their cyberinfrastructure (CI).

In this paper, we present the design of a push-based data delivery framework that leverages emerging in-network capabilities, along with data pre-fetching techniques based on a hybrid data management model. Specifically, we analyze data access traces for two large-scale observatories, Ocean Observatories Initiative (OOI) and Geodetic Facility for the Advancement of Geoscience (GAGE), to identify typical user access patterns and to develop a model that can be used for data pre-fetching. Furthermore, we evaluate our data pre-fetching model and the proposed framework using a simulation of the Virtual Data Collaboratory (VDC) platform that provides in-network data staging and processing capabilities. The results demonstrate that the ability  of the framework to significantly improve data delivery performance and reduce network traffic at the observatories' facilities.

\end{abstract}

\section{Introduction}

Almost all of science is transitioning from ``data-poor'' to ``data-rich'' in the $\rm 21^{st}$ century, and large-scale observatories and shared-use instruments are playing essential roles in catalyzing scientific discoveries. With the increasing availability of shared-use instruments and observatories, observational data is becoming an important part of application workflows and contributor to scientific discoveries in a range of disciplines, and has enabled key scientific discoveries~\cite {abbott2016observation,akiyama2019first}. 

However, the corresponding growth in the number of users accessing these facilities coupled with the expansion in the scale and variety of the data, is making it challenging for these facilities to ensure their data can be accessed, integrated and analyzed in a timely manner, and is resulting in significant demands on their cyberinfrastructure (CI). At the same time, applications that use data from these observatories expect robust, highly available and performant data services with low access latency~\cite{rodero2019data,deelman2019cyberinfrastructure,aaai2020}. Consequently, it is essential to address the associated data management and delivery challenges.

Recent studies have addressed these challenges from different perspectives. Several large-scale facility projects~\cite{rodero2019data,dewdney2009square,abramovici1992ligo,kampe2010neon,deelman2015pegasus,albrecht2012makeflow} co-design their facilities, cyberinfrastructure, and applications to support highly-customized and high-performance end-to-end scientific workflows. Although this supports the integration of processes from data generation to final discovery for its targeted workflows, developing such a specialized cyberinfrastructure is time-consuming and expensive, and the resulting solutions are typically inflexible. 

Existing research has also explored reducing the volume of data transferred using methods such as data streaming~\cite{roderoenabling}, in-transit processing~\cite{zamani2017deadline}, edge processing~\cite{renart2019edge}, continuum computing~\cite{balouek2019towards}, and proxy caching\cite{ali2011survey,zhang2013caching,gagliardi2011content,jiang2018cachalot}. However, these solutions require appropriate capabilities at the edge and/or in the network and a co-design of user applications, limiting their broad adoption.

Recent CI projects~\cite{PRP,parashar2019virtual,altintas2019workflow} equipped with high-speed networks aim to address the data access challenge. For example, the Virtual Data Collaboratory (VDC) leverages the science DMZ network model~\cite{magri2014science,calyam2014wide,farrell2016science} to construct a dedicated regional network to enable high-throughput data transfers among users, data sources, and computing resources. However, considering that observational data is being generated at rates greater than increases in network bandwidth, the advances in  networks alone cannot fundamentally resolve this challenge.

Furthermore, users access observational data in multiple ways with different performance expectations and access patterns. Users can manually download data from the observatory website, run scripts that automatically query data, or subscribe to real-time data streams. 
For example, automated script-based downloads are typically a part of application workflows~\cite{deelman2019cyberinfrastructure,deelman2015pegasus}, and while they often required stricter performance guarantees, their query and access patterns are more predictable. 

We believe that an integration of knowledge of users' data access behaviors with the CI's data-delivery mechanisms can address data access challenges, improving data delivery performance as well as overall CI efficiency. 
In this research, we study user access and data usage patterns by analyzing the access traces of two NSF-funded large-scale observatories, the Ocean Observatories Initiative (OOI) operated by Rutgers and the Geodetic Facility for the Advancement of Geoscience (GAGE) operated by UNAVCO. At the same, we experiment with the VDC platform, a Science DMZ-based data CIs, and explore how its Data Transfer Nodes (DTNs) can be exploited to cache and pre-fetch data. Based on this study and its findings~\cite{qin2019towards}, we develop a set of optimization strategies that include a distributed cache layer, hybrid pre-fetching model (history-based and data mining-based), data placement strategy, and streaming mechanism. Furthermore, we combine these strategies and propose a push-based data delivery framework that can accelerate data delivery performance for large-scale shared-use observatories.
The proposed framework is designed to run on DTNs within the VDC platform between the end-users and the data sources, allowing users to discover and access data through their local DTN, and the actual transfer of data is handled by the network of DTNs that are part of the framework using the optimization strategies developed in this work. From a end-users' perspective, data access is typically local as data is often pre-fetched and cached at the local DTN.

We have evaluated this model and framework using a simulated VDC platform and access traces from OOI and GAGE. Specifically, we measure latency and throughput to quantify the impact on data access performance of end-users under different network conditions, request traffic, and cache configurations. Moreover, we compare our hybrid pre-fetching model with state of the art pre-fetching models~\cite{xiong2016prefetching,li2012prefetching} for spatial-temporal data. The results show that our framework, combined with the proposed optimization strategies, can significantly improve data delivery performance and reduce the load on the observatory's CI. The main contributions of this work are as follows:

\begin{itemize}[]
    \item Analysis of user access and data usage patterns for two representative large-scale shared-use observatories.
    \item Development of a hybrid data management model designed to support pre-fetching mechanisms.
    \item Use of DTN's within a Science DMZ to support data pre-fetching and caching based on user data access patterns.
    \item Design and prototyping of a push-based data delivery framework along with optimization strategies that leverage user data access patterns for data pre-fetching and caching.
    \item Experimentally evaluate the push-based data delivery framework using a simulated VDC platform and  access traces from  OOI and GAGE.
\end{itemize}

The remainder of this paper is organized as follows. 
Section~\ref{Sec:background} presents motivations for the presented work.
Section~\ref{Sec:observatory_usage_analysis} analyzes the user data access patterns using OOI and GAGE access traces. 
Section~\ref{Sec:system_design} describes the design push-based data delivery framework along with optimization strategies for data pre-fetching and caching. 
Section~\ref{Sec:experiment} presents a performance evaluation of the system. 
Section~\ref{Sec:conclusion} concludes the paper.

\section{Motivation and background}
\label{Sec:background}

\begin{figure*}[!htb]
\centering
\includegraphics[width=0.8\textwidth]{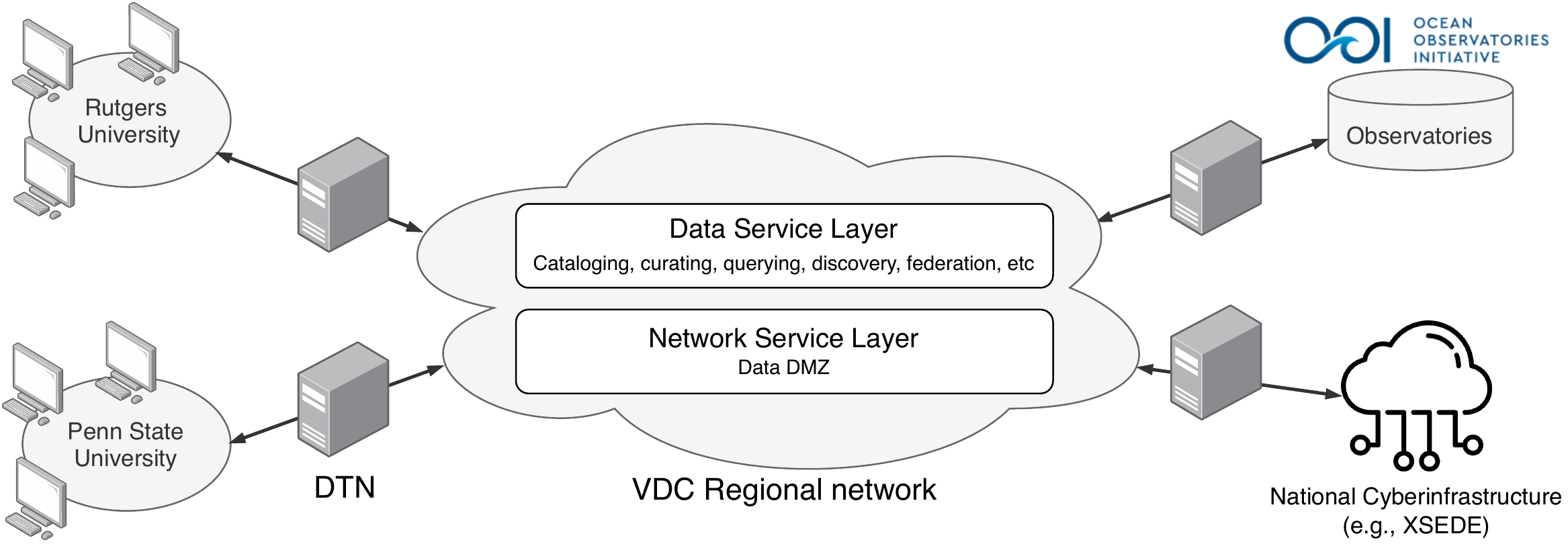}
\caption{Overall architecture of the Virtual Data Collaboratory (VDC). VDC is a data DMZ-based regional network. Users can connect to VDC using Data Transfer Nodes (DTN). The Data Services Layer support data discovery, access and processing across a federated environment that integrates multiple data sources and CI.}
\label{fig:vdc_architecture_2}
\vspace{-0.5cm}
\end{figure*}

\subsection{Observatory data access challenge}
\label{sec:motivation_big_data}

Large-scale, shared-use scientific facilities, such as shared instruments, observatories and experimental platforms, have become key enablers for scientific discoveries in a range of disciplines. For example, the Event Horizon Telescope (EHT) generated the first images of a black hole event horizon~\cite{akiyama2019first}, and the Laser Interferometer Gravitational-Wave Observatory (LIGO) enabled gravitational waves to be detected for the first time~\cite{abbott2016observation}. As a result, ensuring efficient, pervasive and democratic access to data from such facilities is essential and critical to amplifying their scientific impact.

However, rapidly growing data volumes and rates and the lack of sufficiently CI capabilities can limit effective access, and prevent scientists from using the data in a timely manner. For example, the LIGO project generates terabytes of data every day during its ``observing" mode~\cite{ligo}, and the in-construction Square Kilometer Array (SKA) project is estimated to generate an exabyte of raw data every day though it will then be compressed to around 10 petabytes~\cite{ska_web}. As noted at the 2019 NSF Workshops on Connecting Large Facilities and Cyberinfrastructure~\cite{deelman2019cyberinfrastructure,2019_LF}, many large facilities still struggle to provide broad data transfer and access services, especially in remote environments.
Furthermore, using the wide-area network (WAN) is not the most effective approach for transferring large data volumes. For example, Dart et al.~\cite{dart2017assessment} note that it took three months to transfer 56 terabytes of climate model output data from the distributed Coupled Model Intercomparison Project (CMIP5) archive to the National Energy Research Supercomputing Center (NERSC). 
As most facilities do not provide co-located general-use computing and storage resources for end-users, users and application workflows typically have to retrieve the data over the network to local or national resources (such as those provided by Extreme Science and Engineering Discovery Environment (XSEDE)). Therefore, addressing data access challenges for large facilities is critical for both scientific facilities and their users.

\subsection{Cyberinfrastructure and science DMZs}

Cyberinfrastructure (CI) aims to knits together the end-users, data sources, software services, and computational resources using high-performance networks~\cite{2019_ci_blueprint},
with the overarching goal of enabling advanced science and engineering application. As the scale and complexity of applications increase, general-purpose networks quickly become insufficient to support their requirements, which can include managing and transporting of terabyte- or petabyte-scale dataset~\cite{crichigno2018comprehensive}.

The Science Demilitarized Zone (Science DMZ) network model \cite{dart2014science} is developed to optimize the transfer of large-scale scientific data. It achieves this goal by building a high-bandwidth network (e.g., 100 Gbps) among campuses, using dedicated Data Transfer Nodes (DTNs), by-passing traditional campus firewalls, and exploiting other techniques, as discussed in~\cite{calyam2014wide,magri2014science}, to support high-throughput disk to disk data transfer~\cite{farrell2016science}.

Several recent-developed CI use the Science DMZ, such as the Pacific Research Platform (PRP) \cite{PRP} and the Virtual Data Collaboratory (VDC)\cite{parashar2019virtual}. In this work, we leverage VDC, which architecture is presented in Figure~\ref{fig:vdc_architecture_2}. VDC constructs a high-bandwidth regional network connection. Its network services layer adopts the Science DMZ model to allow users on different campuses, data sources, and computational resources to connect to the VDC network using data transfer nodes (DTN). Its data service layer provides services, such as cataloging, curating, and querying, which enables users to discover data across observatories that are part for the VDC and integrate this data into application workflows running on remote computational resources. 

DTNs within the VDC architecture are used to temporarily host data as moves across the network. Physically, the DTN is a purpose-built Linux server with significant storage and non-trivial computing power that is configured with the data transfer services such as Globus' gridFTP~\cite{chard2016globus,allcock2005globus,radic2007optimization}, and optimized for receiving WAN transfers at high speed~\cite{liu2018toward,kettimuthu2018transferring}. For example, a DTN is capable for executing computation-intensive tasks such as machine-learning as proposed by the Chase-CI project~\cite{altintas2019workflow}. In this work, we propose to deploy a cache/pre-fetching layer on the DTN to accelerate the data access performance for data from observatories. We also propose to leverage DTNs for data fusion and in-transit data processing.

\subsection{Proxy caching and pre-fetching}

Proxy caching is a well-known strategy for improving data access performance for users in geographically distributed scenarios. It achieves this goal by keeping content that is likely used in the near future at the locations close to its users~\cite{ali2011survey,zhang2013caching}. Web data pre-fetching is a mechanism for predicting and fetching data prior to future requests~\cite{kroeger1997exploring, pallis2008clustering,huang2008mining}. Caching, combined with the pre-fetching, is an effective strategy for improving the user data access experience.

The proxy caching mechanism is typically implemented at three levels based on the cache location: client level, proxy level, and server level~\cite{ali2011survey}. The proxy level caching is widely used in industry to keep the content close to users but not occupy users' local storage space. The Content Delivery Networks (CDN) ~\cite{gagliardi2011content} is one implementation of proxy level caching.

The role of the DTN within the Science DMZ network is equivalent to a proxy level cache within a web system. Therefore, by deploying a cache layer onto the DTNs in VDC essentially constructs a proxy server network that functions as the CDN to accelerate user data access performance~\cite{mokhtarian2014caching} but without adding any additional hardware.

The cache replacement algorithm is crucial for web caching due to cache storage limitations~\cite{ali2011survey,zhang2013caching,podlipnig2003survey}. Wong et al. categorize commonly used web cache replacement algorithms into: 1) Recency-based (e.g., Least-Recently-Used~\cite{vakali2000lru}); 2) Frequency-based (e.g., Least-Frequently-Used~\cite{cherkasova2001role}); 3) Size-based, which evicts the largest object first~\cite{jin2001greedydual}; and 4) Function-based, which evicts objects according to their utility value~\cite{cao1997cost}. The recency-based algorithm is the most commonly used for web caching because the web data has timeliness properties. Moreover, by leverages machine learning techniques, efforts such Ali et al.~\cite{ali2012intelligent,ali2014performance} further improve the Least-Recently-Used algorithm's performance.

 Pre-fetching mechanisms enhance the caching performance by fetching data prior to the users' requests~\cite{kroeger1997exploring, pallis2008clustering,huang2008mining}. Pre-fetching algorithms are usually categorized as content-based and history-based. Content-based methods predict a user's future requests by analyzing data content. Xu et al.~\cite{xu2004keyword} propose to predict a user's future requests based on semantic preferences of previously retrieved Web documents. The history-based method conducts predictions based on  a user's history records, and is the most commonly used approach for web pre-fetching. For example, existing studies~\cite{xiong2016prefetching,yang2001mining} leverage data mining techniques to learn user access patterns using mining the correlations from a user's access records. Other work~\cite{li2012prefetching, nanopoulos2003data} has used Markov-based approaches to predict future requests by matching the user's current access sequence with history access sequences. 

However, as existing literature~\cite{wong2006web} indicates, the factors determining which data objects will be re-accessed can vary significantly for different situations. Understanding user access patterns and particular environment characteristics are important for designing effective cache replacement and data pre-fetching mechanisms. 

In order to exploit the locality of DTNs and the benefit of caching and pre-fetching, we propose to design a push-based data delivery framework along with a set of optimization strategies, including a hybrid pre-fetching model. To maximize the effectiveness of caching and pre-fetching we start by analyzing the access traces from real-world observatory and learn user data access patterns. We then develop a customized caching and pre-fetching mechanism that leverages this knowledge.

\section{A study of observatory data access and usage patterns}
\label{Sec:observatory_usage_analysis}

In this section, we analyze traces from the NSF Ocean Observatory Initiative (OOI) and the Geodetic Facility for the Advancement of Geoscience (GAGE) with the goal of understanding how users access and utilize the data from  these observatories. 

 OOI~\cite{ rodero2019data,ooi-tos,ooimartech16} is a networked ocean research observatory with arrays of instrumented water column moorings and buoys, profilers, gliders, and autonomous underwater vehicles (AUVs) distributed across different open ocean and coastal regions. OOI provides ocean scientists, educators, and the public the means to collect sustained, time-series datasets to enable the examination of complex, interlinked physical, chemical, biological, and geological processes operating throughout the coastal regions and open ocean.

GAGE~\cite{nsf_gage} (operated by UNAVCO) is a non-profit university-governed consortium that facilitates geoscience research and education using geodesy. GAGE services include the operation of the Network of the Americas (NOTA), which integrates a number of existing networks, and engineering, instrumentation, and data services  for user of terrestrial and satellite geodetic technologies.

The analyses presented below is conducted on two traces: a one-month (November 2018) OOI user access log that contains 17.9 million user requests, and a one-year (2018) GAGE log that contains 77.8 million user requests. Each entry in the logs includes a publicly available IP address and  the request metadata (e.g., filename, access timestamp, etc.).

\subsection{Data transfer volume and network conditions}

\begin{figure}[!htb]
\centering
\includegraphics[width=\columnwidth]{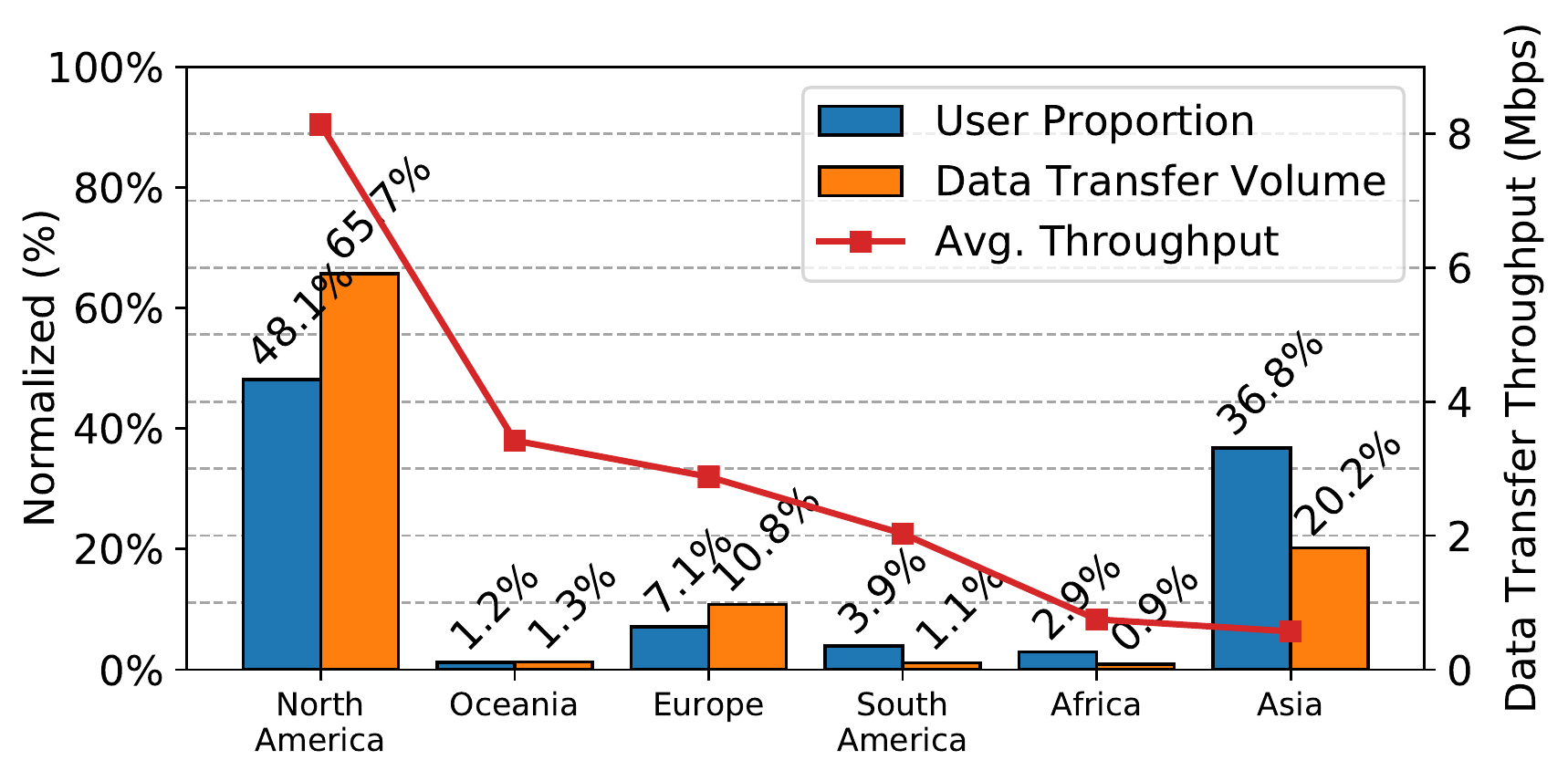}
\caption{Study of the impact of WAN performance on GAGE users' data access from different continents. The plot compares the proportion of users, data transfer volume, and the associated average data transfer throughput for each continent.}
\label{fig:unavco_study_case}
\end{figure}

We use the GAGE trace to study the impact of network performance exacerbates on data access. Since GAGE has a global user community, we analyses access performance across different continents. First, we estimate the user locations by reverse tracking their public IP address. Then, we compute the user distribution, the associated data transfer volume, and the average data transfer throughput as shown in Figure~\ref{fig:unavco_study_case}.  It is worth noting that while GAGE does have users from Antarctica who operate GPS receivers and work with data from the McMurdo station, their access records appear to come from other continents.

We observe that available network speed directly impacts the volume of user data transfers.  North America, Oceania, and Europe have the highest average network throughput, and there is a positive correlation between their user percentages and the associated data transfer volumes. In contrast, these is a negative correlation for the rest of the continents. In particular, for Asian users, although they account for 37\% users, they only account for 20\% of the data transfer volumes and also have the lowest network performance (about 0.568 Mbps). This implies that low network performance limits user data access from the observatory.

\subsection{Classification of users and requests}
\label{sec:user_request_classification}

Data from the observatories is spatial-temporal and contains geographic and timing information in its metadata. To query this data, users need to provide the name of the data object and an observation time range, i.e., the starting and ending timestamp.  

Our analysis of the OOI and GAGE traces reveals two types of requests: \textit{human requests}, which are interactive requests from humans browsing the data catalog and manually downloading data, and \textit{program requests}, which are from workflows or scripts that routinely download data using automated APIs. 

We distinguish between these types of requests based on their access patterns, which differ in terms of their access frequencies and the queried data sets. Human requests query a set of data within a short period and then disappear. In contrast, program requests regularly query the latest data at a consistent frequency. We classify the users corresponding to these request types as \textit{human user} and \textit{program user} respectively.

To identify the request types, we maintain a running time window (e.g., one week) and analyze $(i)$ each user's request frequency and $(ii)$ patterns of repetition in the requests within that time window. If a user requests the same set of data objects more than once per day and this pattern repeats every day during the time window, we classify this user's future requests for this set of data objects as program requests. All other requests are classified as human requests. 

Since access frequencies and the queried datasets differ between human users and program users, they require different optimization strategies. As a result, it is important to understand the relative percentages and associated data access volumes for each user type.

\begin{table}[t]
\centering
\resizebox{0.99\columnwidth}{!}{
\begin{tabular}{rcclcc} 
\toprule
     & \multicolumn{2}{c}{Number of users (\%)} &  & \multicolumn{2}{c}{Data transfer volume (\%)}  \\ 
\cline{2-3}\cline{5-6}
\\[-0.9em]
     & HU     & PU                             &  & HU    & PU                                     \\ 
\hline\hline
\\[-0.9em]
OOI  & 86.7\% & 13.3\%                         &  & 9.9\% & 90.1\%                                 \\
GAGE & 94.1\% & 5.9\%                          &  & 9.4\% & 90.6\%                                 \\
\bottomrule
\end{tabular}
}
\caption{Percentage of \textit{Human Users} (\textbf{HU}) and the \textit{Program Users} (\textbf{PU}), and their data transfer volumes for OOI and GAGE, respectively. }
\label{table:user_population_data_volume_analysis}
\end{table}

\subsection{A quantitative study of users and their data accesses}
\label{sec:user_data_req_usage_classification}

The user percentages and associated data transfer volumes for human users and program users for OOI and GAGE are presented in Table ~\ref{table:user_population_data_volume_analysis}. More than 86.7\% of users are human users for both the observatories making them the primary user type. However, they generated less than 9.9\% of the data transfers, which implies that program users are the primary data consumers. As a result, we mainly focus our optimizations on requests by program users and address human users to the extent possible.

\subsection{Analysis of program requests}
\label{sec:requests_analysis}

\begin{figure}[!htb]
\centering
    \begin{subfigure}[h]{0.49\textwidth}
        \includegraphics[width=\textwidth]{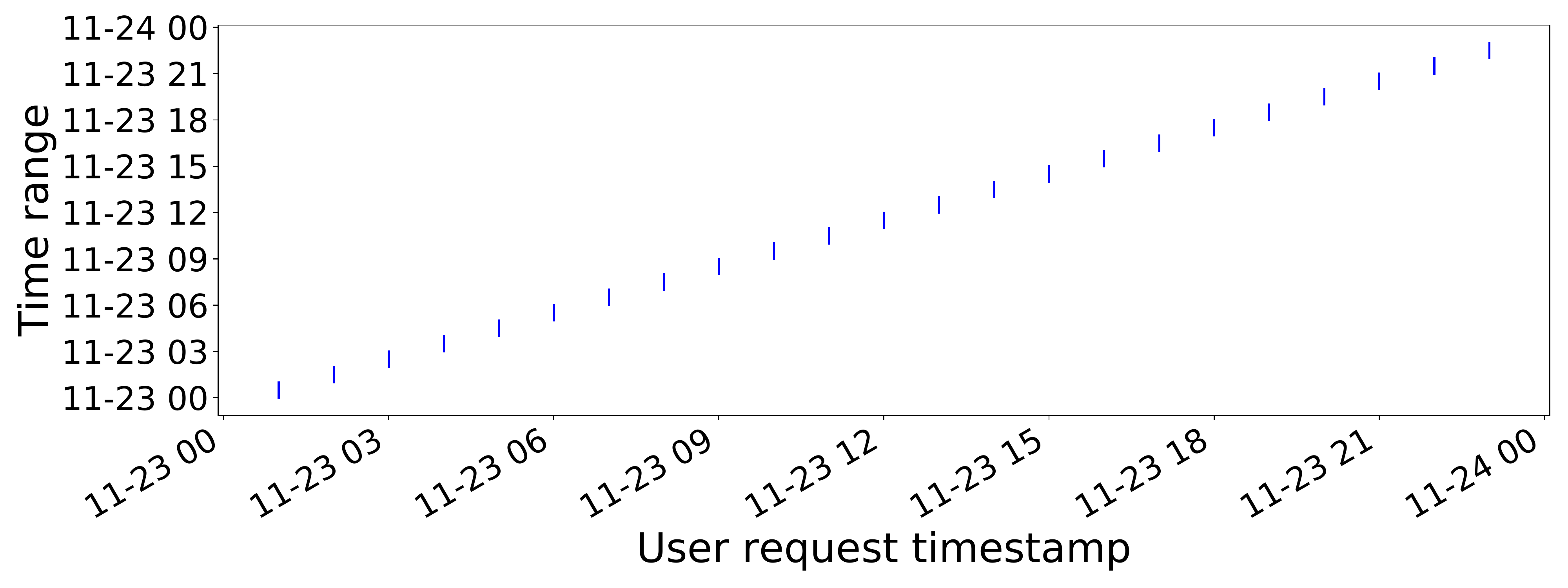}
        \caption{Regular request}
        \label{fig:user_legitimate_request}
    \end{subfigure}
    \begin{subfigure}[h]{0.49\textwidth}
        \includegraphics[width=\textwidth]{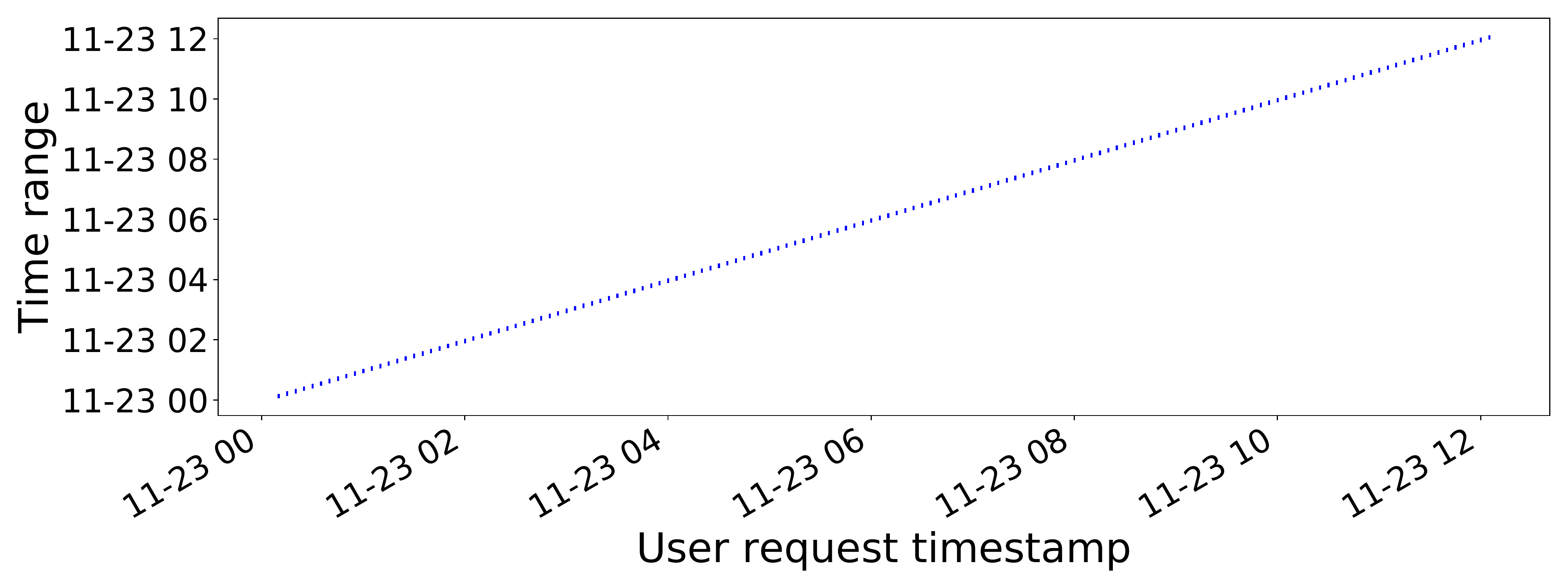}
        \caption{Real-time request}
        \label{fig:user_realtime_request}
    \end{subfigure}
    \begin{subfigure}{0.49\textwidth}
        \includegraphics[width=\textwidth]{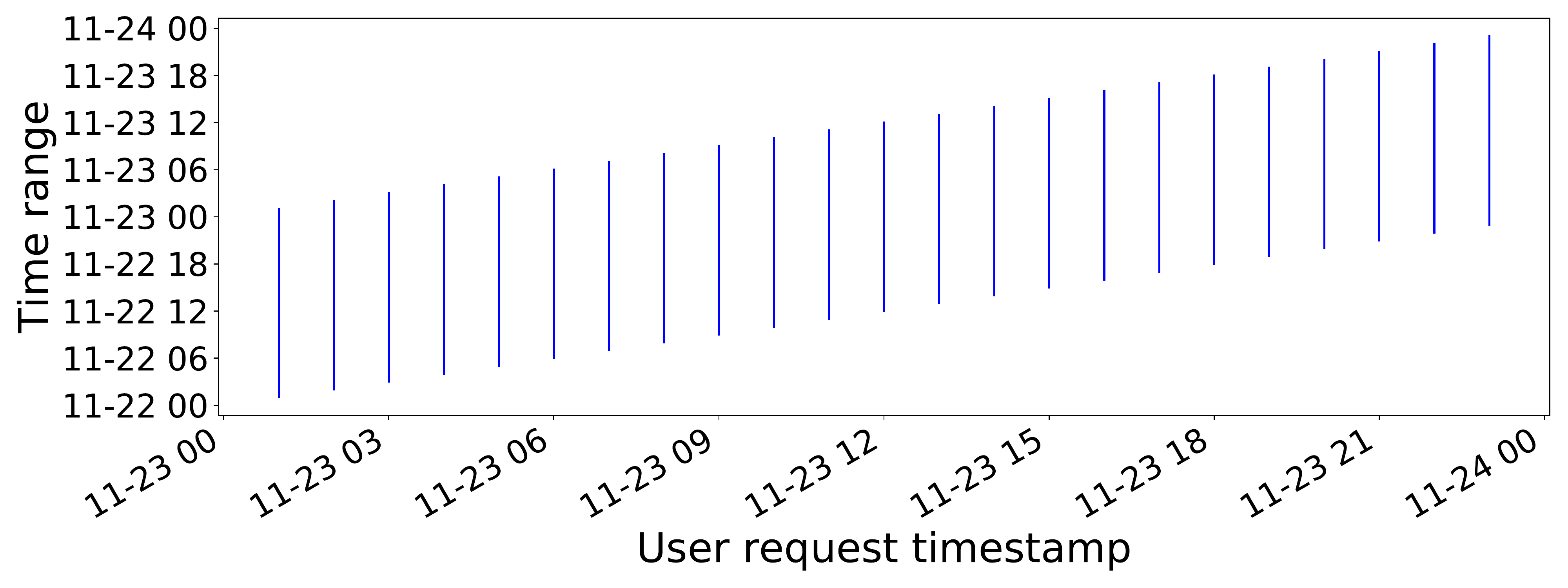}
        \caption{Overlapping request}
        \label{fig:user_faulty_request1}
    \end{subfigure}
    \caption{Examples of the three types of user requests in the OOI trace on 23 November 2018.  Figure~\ref{fig:user_legitimate_request} illustrates \textit{regular} requests where a user requests the past one-hour of data every hour. Figure~\ref{fig:user_realtime_request} illustrates \textit{real-time} requests where a requests the past one-minute of data every minute (only a subset of the requests are plotted for visualization purposes). Figure~\ref{fig:user_faulty_request1} illustrates \textit{overlapping} requests where a user requests the past one-day of data every hour.}
    \label{fig:user_three_types_request}
\end{figure}

We analyze program requests based on the request metadata, i.e., access time, data object, and time range, and identify three access patterns: \textit{regular}, \textit{overlapping} and \textit{real-time} request. These three request types are illustrated in Figure~\ref{fig:user_three_types_request} using a sample from the OOI trace on 23 November 2018. The horizontal axis is the time of the access request, and the vertical axis is the time range requested (i.e., the starting and ending timestamp of the requested data). Each vertical blue bar in the plot is a request, and the length of the bar represents the requested time range.

\textbf{Regular requests} query new data since the last request without any overlap. For example, in Figure~\ref{fig:user_legitimate_request}, a user makes a request every hour, and each request queries the most recent one-hour of data. 

Regular request represent the most common request type. Observatories typically receive new data and update their databases at regular intervals which is determined by the characteristics of the instruments, the CI design, and other factors such as the data acquisition mechanism (e.g., satellite, cruise recovery). As a result, users develop programs that download the most recently updated data at these regular intervals.

\textbf{Real-time requests} are high-frequency (e.g., once per minute) regular requests.
This type of request is shown in Figure~\ref{fig:user_realtime_request}, where a user queries the past one-minute data every minute. 

Real-time or near real-time data monitoring is typically used to develop science gateways~\cite{roderoenabling} that monitor event occurrences, such as earthquake detection~\cite{aaai2020}. Since most observatories do not support subscribe-based data delivery, 
programs use high frequency \textit{pull-based} requests to implement such monitoring. 
Real-time requests can result in large request traffic and significant load at the observatory. Furthermore, the data returned for each real-time request is typically small. As a result, these requests result in a large number of relatively small data transfers, which can potentially increase the system overheads and, in turn, reduce the quality of service provided by the observatory.

\textbf{Overlapping requests} are similar to the regular requests but with overlapping time-intervals across consecutive requests. These types of queries are illustrated in Figure~\ref{fig:user_faulty_request1}, where a user queries the past one-day data every hour, resulting in a 23-hour overlap between consecutive requests. While such queries are convenient from a user's perspective (e.g., the user is running an application every hour that needs the most recent 24 hours of data), they can have a negative impact on the observatory as they result in redundant data transfers and unnecessary resource utilization. 

Table~\ref{table:both_user_request} lists the percentage data transfer resulting from each of the three request types for OOI and GAGE. Note that for OOI overlapping requests dominate (i.e., 60.8\%), while for the GAGE regular request dominate (i.e., 77.2\%). 

\subsection{Analysis of overlapping requests}

To quantitatively analyze the redundant data transfers resulting from overlapping requests, we divide the data transferred into  \textit{fresh data}, which is the percentage of data that is not part of previous request, and  \textit{duplicate data}, which is the percentage of redundant data. 

Table~\ref{table:both_user_request} shows that 90.4\% of data transferred in response to overlapping requests is redundant data for OOI. Considering that 60.8\% of OOI requests are overlapping, there is a significant amount of redundant data transfer observed in the OOI log.

Considering there is a large amount of redundant data transfer, adding a cache layer into the CI is an effective approach for reducing traffic and resource requirements at the observatory and improving performance. Furthermore. since a large percentage of the requests have predictable patterns, pre-fetching mechanisms can also be effective.

\begin{table}[t]
\centering
\resizebox{\columnwidth}{!}{
\begin{tabular}{rccclcc} 
\toprule
\multirow{3}{*}{} & \multicolumn{3}{c}{Data transfer volume}   &  & \multicolumn{2}{c}{Data transfer volume}  \\
                  & \multicolumn{3}{c}{for the three request types} &  & \multicolumn{2}{c}{for overlapping requests}  \\ 
\cline{2-4}\cline{6-7}
\\[-0.9em]
                  & Regular & Real-time & Overlapping          &  & Fresh data & Duplicate data                 \\ 
\hline\hline
\\[-0.9em]
OOI               & 13.8\% & 25.7\%   & 60.8\%              &  & 9.6\%     & 90.4\%                        \\
GAGE              & 77.2\% & 6.1\%    & 17.2\%              &  & 10.5\%    & 89.6\%                        \\
\bottomrule
\end{tabular}
}
\caption{Data transfer volume for the three types of requests; Breakdown of the data transferred for overlapping request.}
\label{table:both_user_request}
\end{table}

\subsection{Analyzing correlations across requests}

\begin{figure}[!t]
\centering
    \includegraphics[width=0.9\columnwidth]{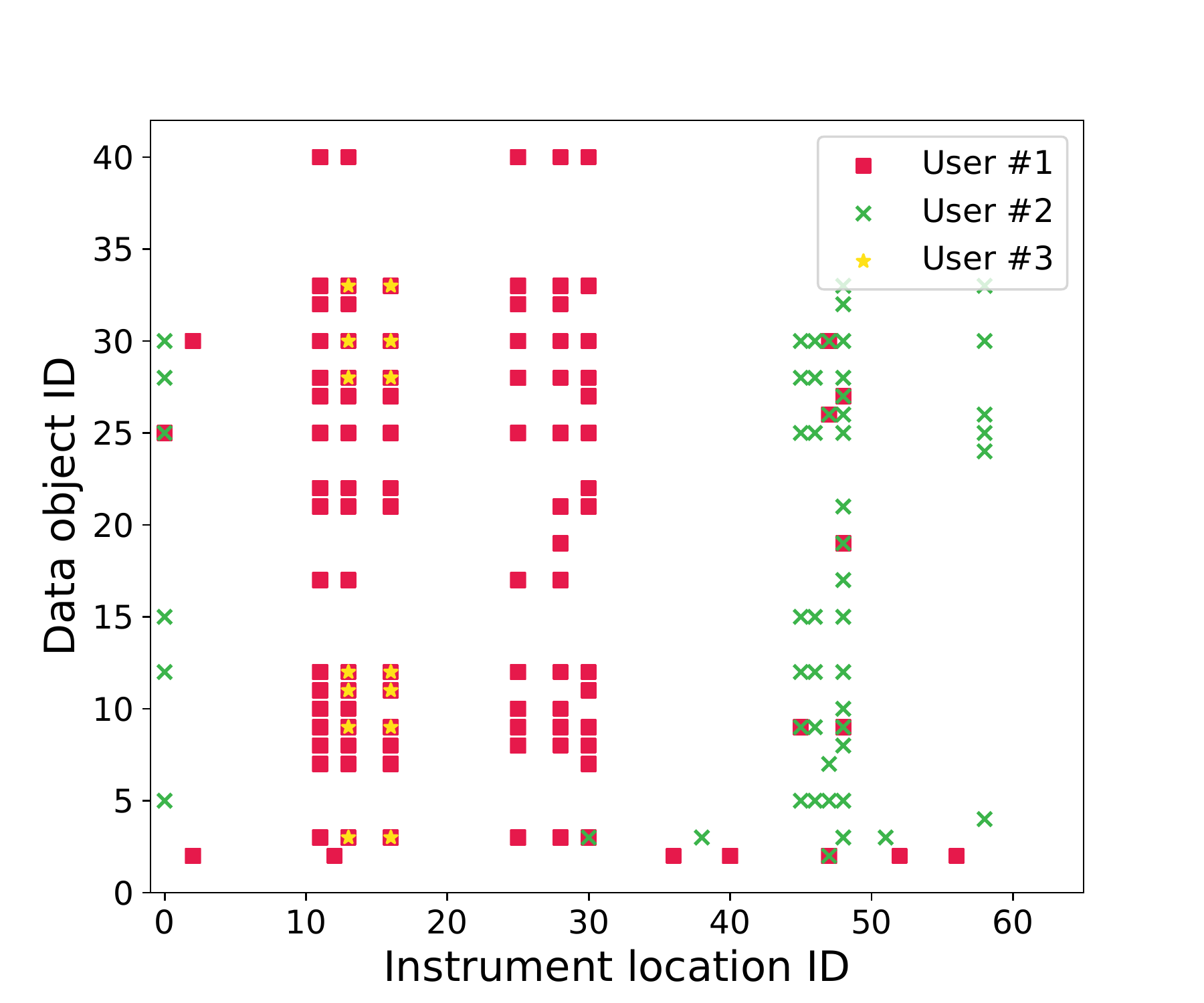}
    \caption{Analysis of user request patterns. The x-axis represents instrument locations, and the y-axis represents requested data object IDs. The visible patterns suggest the existence of spatial correlation across the requests.}
\label{fig:user_request}
\end{figure}

We observe a spatial-temporal correlation across human user requests in the OOI and GAGE traces indicating that human users tend to query data objects in a particular region as they navigate an observatory's data products. There are a couple of potential explanations for this. Scientist may be studying phenomena in a particular regions and as a result, query data from instruments and sensors in that region. Furthermore, observatory data portals typically use maps to help users navigate the observatory and explore desired instruments/sensors and associated data product and as a result, user request tend to spatially correlated. Consequently, it is possible to leverage this spatial-temporal correlation to predict human user requests.

Figure~\ref{fig:user_request} shows the queried data objects (using their instrument name and the instrument location) by three selected users from the one-day OOI trace used in Section~\ref{sec:requests_analysis}. Since OOI deploys the same type of instrument at multiple locations~\cite{ooi_2020}, we use these two pieces of information to represent a specific data object). 
In the figure, we serialize the instrument name into an ID, which is represented in the y-axis. We then serialize the instrument locations, sort them by their proximity, and use their IDs to plot the x-axis. Thus, the dots in the same row represent the same type of instrument at different locations, while the dots in the same column represent different instruments in the same location.

We observe a clear spatial correlation across requests in the plot, as users query multiple data objects of one region (vertical) and the same type of data object in nearby regions (horizontal). We also observe temporal correlation in our analysis. For example, in Figures~\ref{fig:user_three_types_request} where the consecutive blue bars, which represent the time interval of a request, have the same length. This implies that user (especially program users) tend to use a moving window while querying data. These observations confirm that we can potentially leverage this spatial-temporal correlation to predict user requests.

\section{Design of a push-based data delivery framework}
\label{Sec:system_design}

The proposed push-based data delivery framework exploits DTNs to cache queried data, manage the cached data, learn user access patterns, and proactively push data toward appropriate users.

\begin{figure*}[!htb]
\centering
\includegraphics[width=0.95\textwidth]{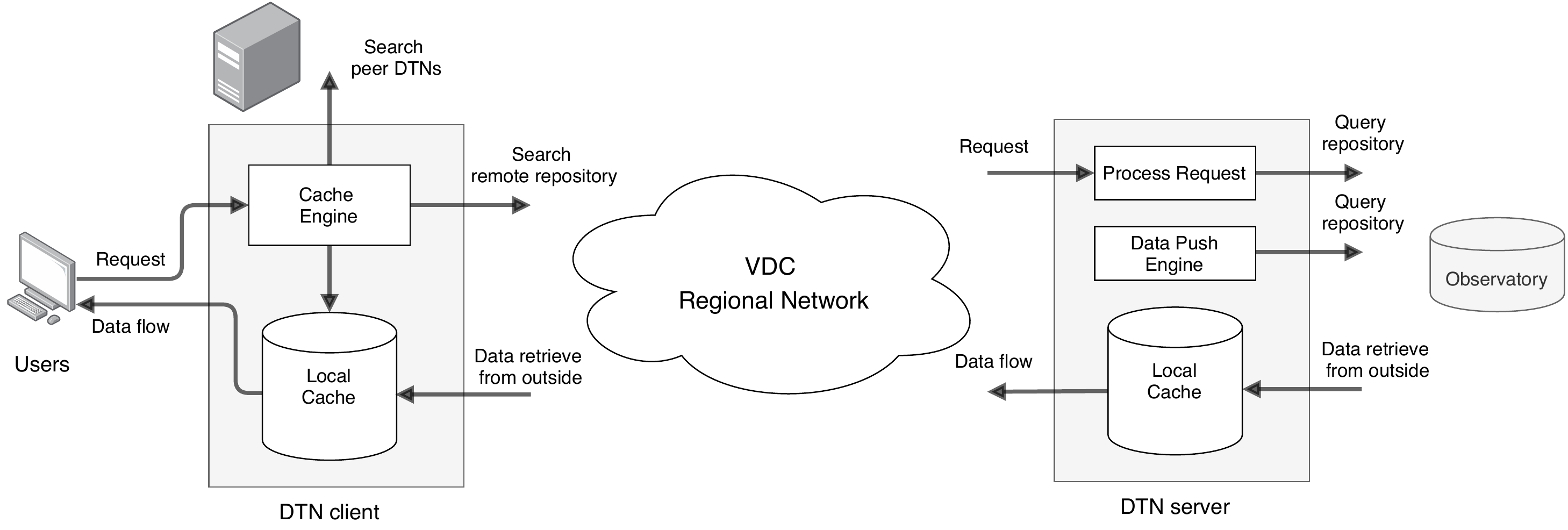}
\caption{Push-based data delivery framework deployed on top of the VDC Science DMZ-base architecture. The framework leverages the DTNs and their storage to implement a distributed and interconnected cache layer. The DTNs also host the data push engine. The framework client runs at the user's local DTNs and handles the user requests. The framework server runs at DTNs at the observatory and is responsible for managing the distributed data cache layer for data pre-fetching and streaming.}
\label{fig:framework_architecture}
\vspace{-0.5cm}
\end{figure*}

The framework has a server-client architecture, as shown in Figure \ref{fig:framework_architecture}. Clients run at user-side DTNs,  pre-process user requests by searching for the requested data in caches at the local and peer DTNs, forwarding the request to the server if data is not found in these caches. The server runs at server-side DTNs that are local to the observatory. It handles requests arriving at the observatory, runs, executes the data pre-fetching and streaming mechanisms, and manages the placement of cached data.

Functionally, the architecture is composed of two main components: the cache layer and the data push engine. The cache layer spans the client and server side DTN to create a distributed interconnected cache network, and uses storage at the DTNs and caches data generated in response to user requests. 
The data push engine hosts the data pre-fetching and streaming mechanisms. It predicts user requests based on access patterns and can pre-fetch or stream data to users according based on these predictions.

\subsection{Design of the hybrid data pre-fetching model}
\label{sec:hybrid_prefetch_model}

The goal of pre-fetching in this work is to predict user accesses based on observed access patterns presented in Section~\ref{Sec:observatory_usage_analysis}, allowing appropriate data objects to be pushed to the user. This is achieved as follows. 

First, we exploit the spatial-temporal correlations to design an \textit{association rule mining-based prediction model} for predicting the human requests as well as other requests that cannot be classified, based on the work from Xiong et al.~\cite{xiong2016prefetching}. Next, we design a \textit{history-based prediction model} to handle program requests using the frequency of these requests. Finally, we develop a data streaming mechanism to process real-time requests.

We use the Autoregressive Integrated Moving Average (ARIMA) model, a widely used technique for time series prediction~\cite{contreras2003arima,faruk2010hybrid}, to predict the timestamp of the user's next request. This combination of models can optimize the prediction accuracy and pre-fetching performance.

\subsubsection{Representing user requests}

We model user requests as a sequence $R_i = \left \langle r_1, r_2,\ldots, r_n\right \rangle$. Each request tuple $r_i$ includes the request timestamp \textit{ts}, the name of the data object (or data stream) \textit{d}, and the requested time range (i.e., the vertical bar in Figure~\ref{fig:user_three_types_request}) \textit{tr}. 
Furthermore, we represent the request of user $j$ as the tuple $R_j$ in  equation~\ref{Eq:request_attribute}, where $TS_n = \left \langle ts_1, ts_2,\ldots, ts_n\right \rangle$, $D_n = \left \langle d_1, d_2,\ldots, d_n\right \rangle$, $TR_n = \left \langle tr_1, tr_2,\ldots, tr_n\right \rangle$, $n$ is the number of data objects queried by the user $j$:

\vspace{-0.5em}
\begin{align}\label{Eq:request_attribute}
  R_j &= \left \langle r_1, r_2,\ldots, r_n\right \rangle \\
  &= \left \langle(ts_1, d_1, tr_1), (ts_2, d_2, tr_2),\ldots,(ts_n, d_n, tr_n)\right \rangle \nonumber \\ 
  &= TS_n, D_n, TR_n \nonumber
\end{align}

\subsubsection{History-based prediction}
\label{sec:historical_based_model}

The history-based data pre-fetching model is used when a request from a program user is identified. Specifically, during the learning period, we monitor a user's request sequence $R$. If its request pattern repeats more than a threshold number of times, then we identify this user as a program user and  mark this series of requests as predictable, and the framework starts pre-fetching data for the user in anticipation of future requests. The learning period and threshold are empirically set to one week and 3 times, respectively in our experiments.

We use the ARIMA model to predict the timestamp of the user's next request, i.e., $ts_{i+1}$. Our implementation of the model is based on existing literature~\cite{contreras2003arima,faruk2010hybrid}. Training ARIMA requires the $n$ most recent data points (i.e., a historical time interval). The higher the value of $n$ is, the more accurate the prediction result is, but the training time is longer. In our experimental evaluation, we empirically set $n$ to 60, which results in a training time of a few seconds and acceptable accuracy. Since the access intervals for \textit{regular request's} are typically in the order of hours, this ARIMA prediction cost is acceptable.

Once the next timestamp of the user's next request, $ts_{i+1}$, is predicted, the framework can start pre-fetching relevant data. We use a pre-fetching offset to determine how far in advance this data should be pre-fetched. For instance, given offset to 0.8, the framework pre-fetches the data at a timestamp at $ts_i + 0.8*(ts_{i+1} - ts_i)$, where $ts_i$ is the timestamp of the user's last request.

The pre-fetching offset allows the user to control the pre-fetched timestamp and allows for time to transfer the data. Furthermore, ARIMA predictions are impacted by the considerable variance across consecutive access times and it can take several cycles for ARIMA to adapt, and an appropriate offset provides sufficient time to buffer the ARIMA prediction latency and achieve a higher pre-fetching success rate.

\subsubsection{Association rule mining-based prediction}
\label{sec:associcated_model}

The association rule mining-based prediction model is used for request from human users and in cases where requests  do not have identified patterns.
This model uses the association rule mining Frequent Pattern Growth (FP-Growth) algorithm~\cite{han2000mining} to construct the prediction model similar to existing work~\cite{xiong2016prefetching,pan2017enhanced,li2017replication, xiong2018replication}. 
FP-Growth will find an association among past requests to predict future requests for data objects, i.e., $d_{i+1}$. The time step for the next request is estimated based on the last two requests as follows: $ts_{i+1} = ts_i + (ts_i - ts_{i-1})$ and $tr_{i+1} = tr_i$. 

The overall model is constructed as follows:

\textit{a)} FP-tree construction: The frequent-pattern tree (FP-tree) is a compact structure that stores quantitative information about frequent patterns in a database. The algorithm first scans the training dataset and counts the number of times a data point $d_i \in D_i$, $l_i \in L_i$ appears in the sequence of requests. This number is called \textit{support}. Then, the algorithm finds the frequency 1-itemsets by comparing the itemset support with a predefined threshold. Finally, it re-scans the dataset and constructs the FP-tree.

\textit{b)} FP-Growth: Based on the FP-tree, which contains association rules between item sets and their corresponding \textit{confidence}, the algorithm filters out association rules that have a confidence value lower than a predefined threshold, to form the complete set of frequent patterns.  

Selecting appropriate threshold values for \textit{support} and \textit{confidence} is important to the model performance. In this work, we empirically set them to 30 and 0.5, respectively.

This method's prediction accuracy is lower than the history-based method due to the randomness in human user requests as observed in our experiments. Therefore, once the model generates a list of related data objects for a user sorted by their probability of repetition, the model only pre-fetches the first $n$ data objects within the time range $tr$ that are identical to the user's last request. The choice of $n$ represents a trade-off between moving more data objects to increases the success rate and the resulting larger overheads. We empirically set $n$ to 3 in our experiments.

\subsection{Design of the data streaming mechanism}
\label{sec:streaming_mechanism}

Real-time (or near real-time) data accesses to OOI and GAGE (we observed 25.70\% and 6.10\% real-time requests to OOI and GAGE respectively) are implemented as a sequence of high frequency requests, which results in significant  traffic at the facilities. Like most such scientific observatories \cite{rodero2019data}, these facilities do not support modern push-based models for delivering real-time streaming data. Furthermore, adding native support for data streaming can be a significant investment~\cite{2019_LF}.

In this effort, we explore how these request for real-time (near-real-time) data can be supported more efficiently using in-network (using DTNs) resources, request prediction and data pre-fetching, and design a data streaming mechanism as part of the push-based data-delivery framework. This is done as follows: Once an incoming request is identified as a real-time request, the framework checks the observatory for data availability and pushes the most recent data back to the user for the duration of the request. Furthermore, the same request coming from multiple users can be combined into a single request to the observatory and redundant requests can be filtered out. Finally, this data can be cached at local DTNs to satisfy subsequent requests. 

\subsection{Design of the cache layer}

The cache layer of the push-based data delivery framework designed to leverage storage at the DTNs to place data from the observatories as close to the user as possible. The design is consists of the choice of the cache eviction policy at each of the local DTNs, and an overall data placement strategy across the network of DTNs that is aware of their geographical distribution.

\subsubsection{Choice of the cache eviction policy}

The cache eviction policy at the local DTNs is based on the user access patterns. As observed in the analysis presented in Section~\ref{Sec:observatory_usage_analysis}, over 90.1\% data transfer volume came from program requests that typically use a moving window to query data, and there is an overlap between consecutive requests.

Based on these observations, the least recently used (LRU) can be an effective cache eviction policy for thelocal DTN, which typically  have a relatively large storage capacity (typical DTN configurations have tens to hundreds of GBs of memory and TBs of fast storage). Note that the LRU eviction policy can be replaced by other eviction policies as needed, and we compare the effectiveness of different policies in our experimental evaluation.

\subsubsection{Data placement strategy}
\label{sec:data_placement}

The  goal of the data placement strategy is to place the data (and its replicas) at local DTNs that are close to potential users, and to keep the hot data, i.e., data with a high chance to be re-accessed, in the cache network as long as possible. The overall data placement strategy is composed of \textit{virtual groups} and \textit{local data hubs}.

A virtual group is a group of users that have common data interests and are geographically close to each other. We can place data objects of interest to a virtual group at a DTN that has the best network connective to the corresponding set of users. 

In order to identify virtual groups with common data interests, we use K-Means to cluster past requests and then group users corresponding to the resulting clusters based on geographical proximity. We then map these user groups to appropriate DTNs which serve as the local data hub for the group. Note that this design assumes that users consistently access the observatory using  institutional (e.g., university or research laboratory) resources and that the DTNs within the VDC align with these access locations.

The selection of a local data hub for a virtual group is based on three factors: network throughput, resource availability (e.g. storage) and user  request frequency. As illustrated in equation~\ref{Eq:data_hub_selection}, we select the DTN $V_{dh}$ that maximizes the weighted sum of the three values as the local data hub within the virtual group. In this equation, $P_{ij}$ represents the network throughput from DTN $v_i$ to $v_j$ and $\theta_p$ is the weight for the throughput; $U_i$ represents the DTN device resource utilization (e.g., storage and CPU) with weight $\theta_u$; and $F_i$ represents the request frequency of requests from the virtual group members that are connected to this DTN and has weight $\theta_f$. We empirically set $\theta_p=0.6$, $\theta_u=0.2$, and $\theta_f=0.2$.

\begin{equation}
\resizebox{0.9\hsize}{!}{%
  $V_{dh} = 
    max( \theta_p \sum_{j \neq i}^n P_{ij} + \theta_u U_i + \theta_f F_i ), \ 0 < i,j \leq n$
    }
  \label{Eq:data_hub_selection}
\end{equation}

Note that clustering virtual groups and the selection of local data hubs are performed periodically allowing the framework adapting to users changing their groups and interests. If the local data hub changes, the previous data hub keeps the data that was already cached, and new data is cached at the new data hub to minimize reconfiguration costs.

\begin{figure}[!t]
    \centering
    \includegraphics[width=0.85\columnwidth]{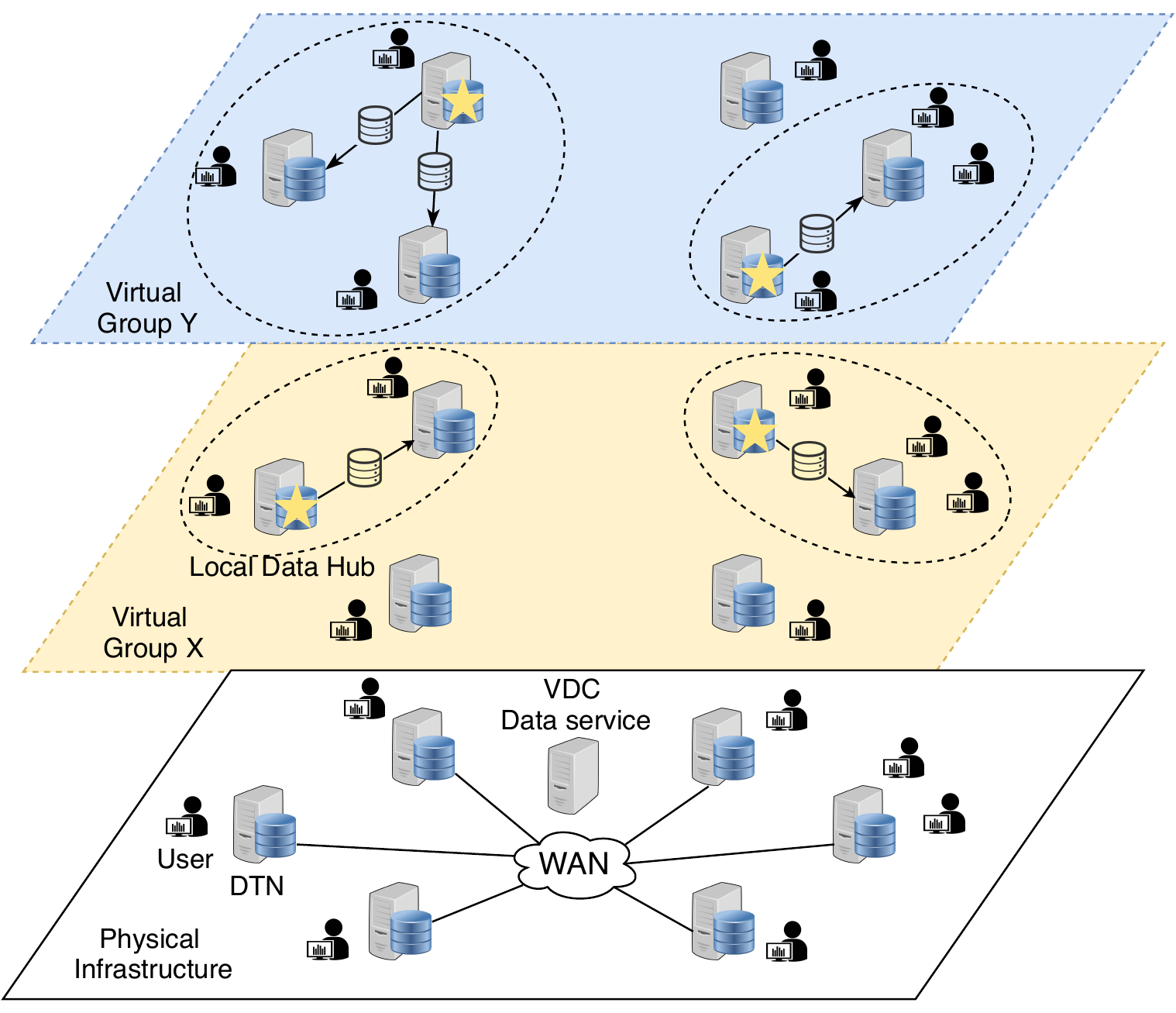}
     \caption{Virtual groups and corresponding local data hub (i.e., the DTN with a star). The layered architecture is overlayed on top of the physical infrastructure. Users can belong to several virtual groups based on their data interests. These groups are split into sub-groups base on their geographic locations into sub-groups (represented by circles).}
    \label{fig:DTN_virtual_group}
\end{figure}

The virtual group concept is illustrated in Figure~\ref{fig:DTN_virtual_group} and is inspired by~\cite{jiang2018cachalot}. As shown in the figure, the virtual groups are overlayed on the physical infrastructure composed of DTNs that serve as the users' access points. Users below on to one or more virtual groups (e.g., \textit{Virtual Group X} and \textit{Virtual Group Y} in the figure) based on their data interests, and each virtual group is divided into sub groups based on their geographical location and mapped to DTNs in the physical infrastructure, which is the local data hub for the sub group serving as its access point and local cache.

\subsection{Overall operation of the framework}

The overall operation of the framework is illustrated in Figure~\ref{fig:framework_architecture}. When the client DTN receives a request, it first searches the cache layer starting with cache at the user's local DTN. Any portion of the requested data found at the local cache is returned, the peer DTNs are searched for any remaining requested data. If data is found cached at peer DTN, it compares data transfer costs from that DTN to transfer costs from the observatory and decided whether to transfer data from the DTN. Any remaining data is then requested from the server, and is also recorded by the pre-fetching engine allowing it to update its models. Finally, the data is pushed towards the user.


\section{Experimental evaluation}
\label{Sec:experiment}

\subsection{Methodology and setup}

We use a simulations along with OOI and GAGE request traces to evaluate the effectiveness and performance of the push-based data delivery framework presented in this paper under different conditions, and quantitatively compare it with two state-of-the-art pre-fetching models~\cite{li2012prefetching,xiong2016prefetching}.

\subsubsection{The Simulator}

\begin{figure}[t]
	\begin{centering}
	\includegraphics[width=\columnwidth]{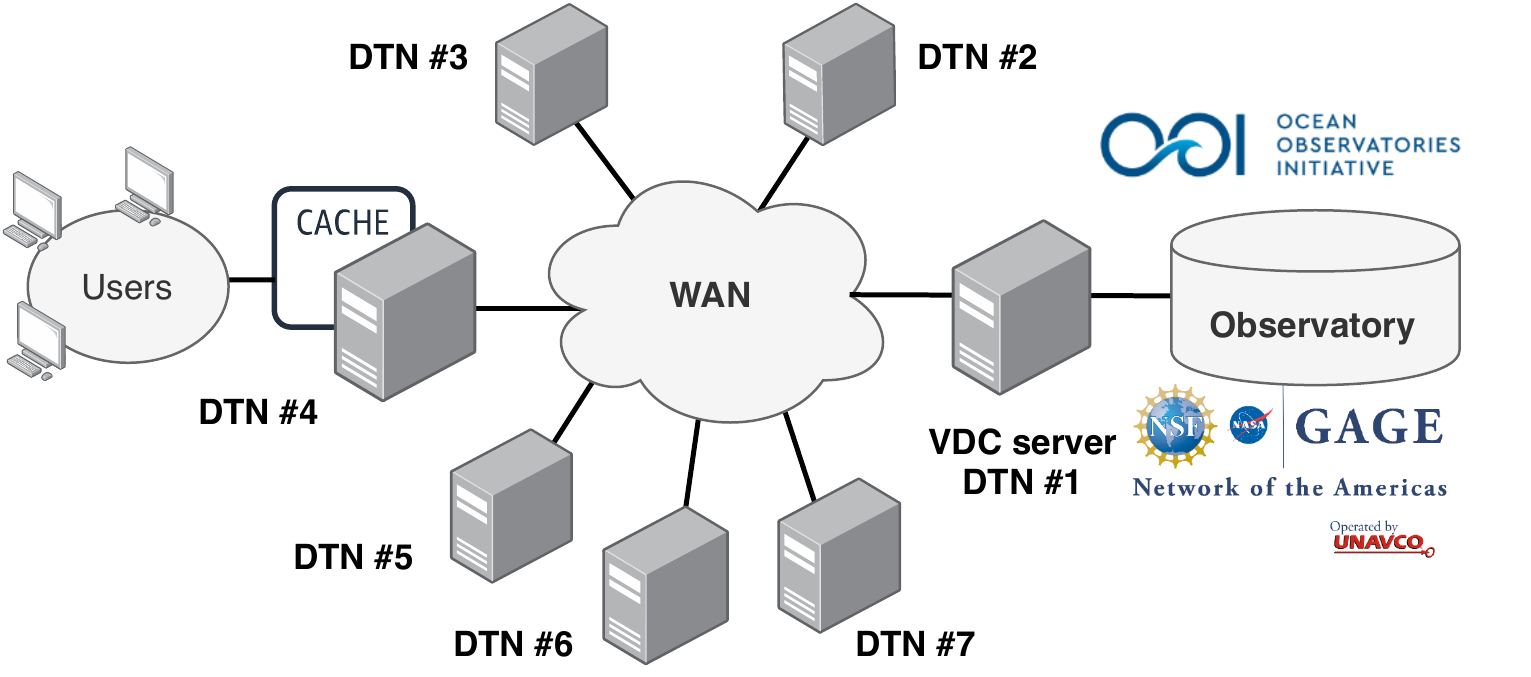}
	\caption{Architecture of the simulator emulating the VDC cyberinfrastructure and used to implement push-based data delivery framework. DTN\#1 as the VDC server and  hosts the pre-fetching engine and manage data placement. The other DTNs (i.e., DTN\#2 - \#7) are client DTNs and collectively form the cache layer.}
	\label{fig:exp_test_bed}
	\end{centering}
\end{figure}

We developed a simulator to emulate the VDC cyberinfrastructure as shown in Figure~\ref{fig:exp_test_bed}. The simulated architecture is composed of seven geographically distributed DTNs interconnected via a wide-area network (WAN). We set DTN\#1 as the VDC server that provides key data services including data cataloging, discovery, querying, etc.

The presented data delivery framework is deployed on the simulated VDC platform. The VDC server DTN is also server DTN for the framework and is the access point for the observatory. This DTN hosts the pre-fetching engine and manage data placement. The other DTNs (i.e., DTN\#2 - \#7) are the framework's client DTNs and collectively form the cache layer. The client DTNs can be configured to run different cache policies and use LRU by default. In our experiment we also evaluate the Least-Frequently Used (LFU) policy.

\begin{figure}[t]
	\begin{centering}
	\includegraphics[width=0.4\textwidth]{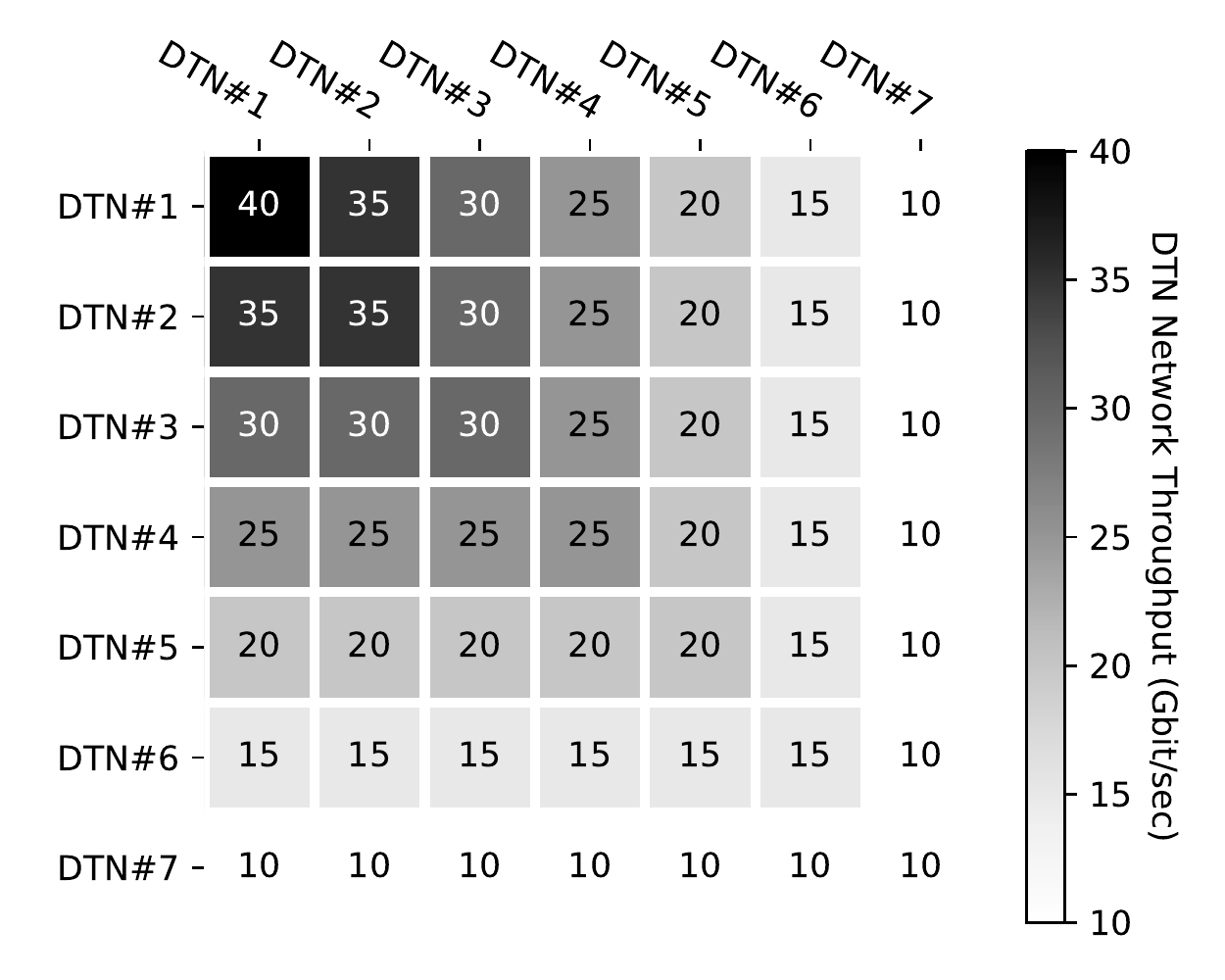}
	\caption{Configuration of the interconnection bandwidths between DTNs used by the simulator.}
	\label{fig:exp_dtn_throughput}
	\end{centering}
\end{figure}

To emulate GAGE's average network throughput shown in Figure~\ref{fig:unavco_study_case}, we limit the client DTNs' bandwidth from 40Gbps to 10Gbps. The network bandwidth between DTNs is shown in Figure~\ref{fig:exp_dtn_throughput}. Since the bandwidth is not homogeneous across DTNs, we also evaluate the impact of the  data placement strategy and the  on the virtual group and local data hub schemes. The simulation assumes that users connect to their local DTNs at 100 Gbps.

The simulation uses a task queue at the VDC server to process the user requests, and ten service processes. User requests arriving at the server DTN are queued in the task queue and wait for the next available service process. Limiting the service processes to ten allows us to evaluate the impact of increasing request traffic on the observatory data service. When user requests arrive faster than the service processes can process, it results in longer queuing time and larger processing latency.

\subsubsection{Evaluating the hybrid pre-fetching model}

We evaluate the Hybrid Pre-fetching Model (referred to as \textbf{HPM}) used by the push-based data delivery framework by comparing it to two state-of-the-art spatial-temporal pre-fetching models~\cite{xiong2016prefetching,li2012prefetching}.
The first reference model (\textbf{MD1}) is by Li et al.~\cite{li2012prefetching} and uses the Markov method to perform predictions. The authors connect geospatial data coordinates to convert the user access history into an ``access path''. They observe that such paths follow Zipf's law, and thus, they can predict user requests using a basic Markov model. In our evaluation, we add the geospatial coordinate information to the OOI and GAGE datasets and use this model.

The second reference model (\textbf{MD2}) is by Xiong et al.~\cite{xiong2016prefetching} and is a data mining-based method that uses a regional mesh and association rules to learn the spatial correlation, and use ARIMA to predict the temporal correlation. This model applies the same prediction strategy to all user requests. In contrast, HPM distinguishes between request types (i.e., human or program requests) and uses the appropriate model for each request type.

\subsubsection{Network conditions and request traffic variations}
\label{sec:variations}

To evaluate data delivery performance under different network conditions, we consider three scenarios: \textit{best}, \textit{medium}, and \textit{worst}. The \textit{best} case maintains the original DTN bandwidth, as shown in Figure~\ref{fig:exp_dtn_throughput}. The \textit{medium} and \textit{worst} cases cap the bandwidth at each DTN at 50\% and 1\%, respectively. 

Request traffic represents the number of requests that the observatory receives within a unit of time. Due to limited computational capacity and bandwidth, the observatory can only process a limited number of requests concurrently, and as a result, as the request traffic increases, the number of pending requests in the task queue also increases, and latency and data transfer times are correspondingly higher. This impact the performance of the data delivery framework.

In our simulations, we consider three request traffic scenarios: \textit{low}, \textit{regular}, and \textit{heavy}. We define the traffic corresponding to the one month traces from the OOI and GAGE as the regular request traffic, and emulate different request traffic scenarios relative to this traffic. To emulate the heavy request traffic, we compress  one month of requests into a one-week time interval. It means that the observatory will receive four times the number requests within a unit time as compared to regular request traffic. Similarly, we emulate low request traffic by expanding one month of request to span a two-month time interval.

\subsubsection{Simulator configurations}
\label{sec:cache_configuration}

As noted previously, the simulator is configured to use the LRU cache eviction policy by default, and our evaluations compares LRU with LFU.
We also compare the performance for different cache sizes. Specifically, due to the difference in data sizes, we evaluate the OOI trace with cache sizes of 128GB, 256GB, 512GB, 1TB, and 10TB, and the GAGE trace with cache sizes of 32GB, 64GB, 128GB, 256GB, and 10TB. Note that the 10TB cache is large enough in enough to cache the entire data and represents best-case performance. 

Given that OOI and GAGE requested are distributed across the globe, in our simulations, we cluster request based on the continent they originate from. We use the client DTNs (DTN\#2 - \#7) to represent the six continents (excluding Antarctica) and distribute users across these DTNs according to their locations. By default, the simulator uses the \textit{best} network configuration and \textit{regular} request traffic configuration.

\subsubsection{Evaluation metrics}

In our experiments, we measure \textit{latency} and \textit{throughput} to quantify the data delivery performance to end-users. \textit{Latency} is defined as the time between the when  the user submits a request that when the observatory starts processing it, and includes the time it spends in the observatory task queue. We compute \textit{throughput}, by dividing the data size by the total data transfer time.

Furthermore, we use the \textit{recall} metric to evaluate the performance of the pre-fetching mechanism, which is the the percentage of the pre-fetched data that is accessed by the user. A higher \textit{recall} value indicates that a smaller amount of pre-fetched data is wasted. Pre-fetched data may be wasteful if the prediction was incorrect, if the data not delivered on time (i.e., before the user requests it), or if the cache configuration caused it to be evicted before it was accessed.

\begin{figure*}[!th]
\centering
\begin{subfigure}{0.3\textwidth}
	\includegraphics[width=\textwidth]{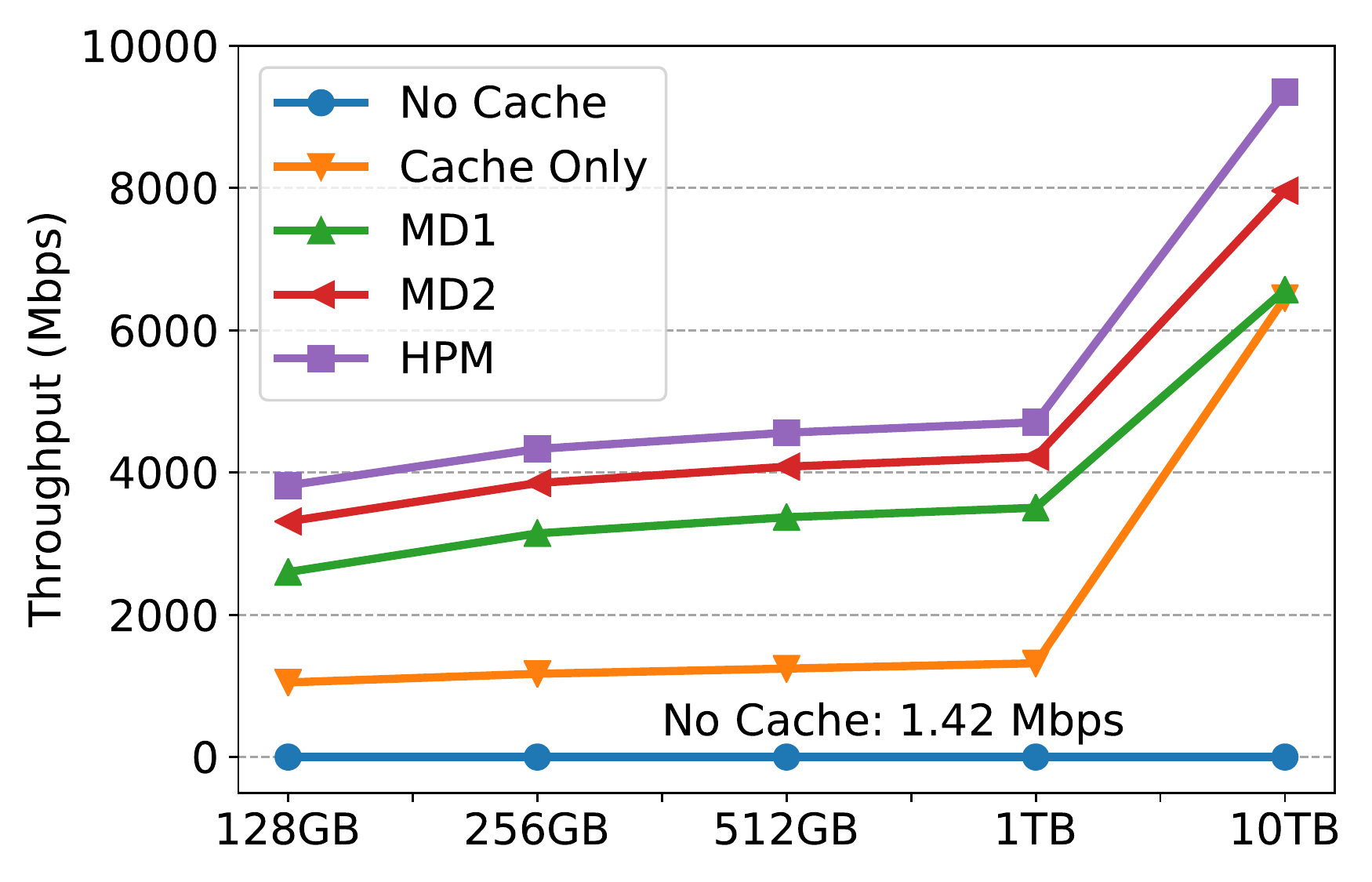}
	\caption{Throughput}
	\label{fig:ooi_cache_throughput_lru}
\end{subfigure}
\begin{subfigure}{0.3\textwidth}
	\includegraphics[width=\textwidth]{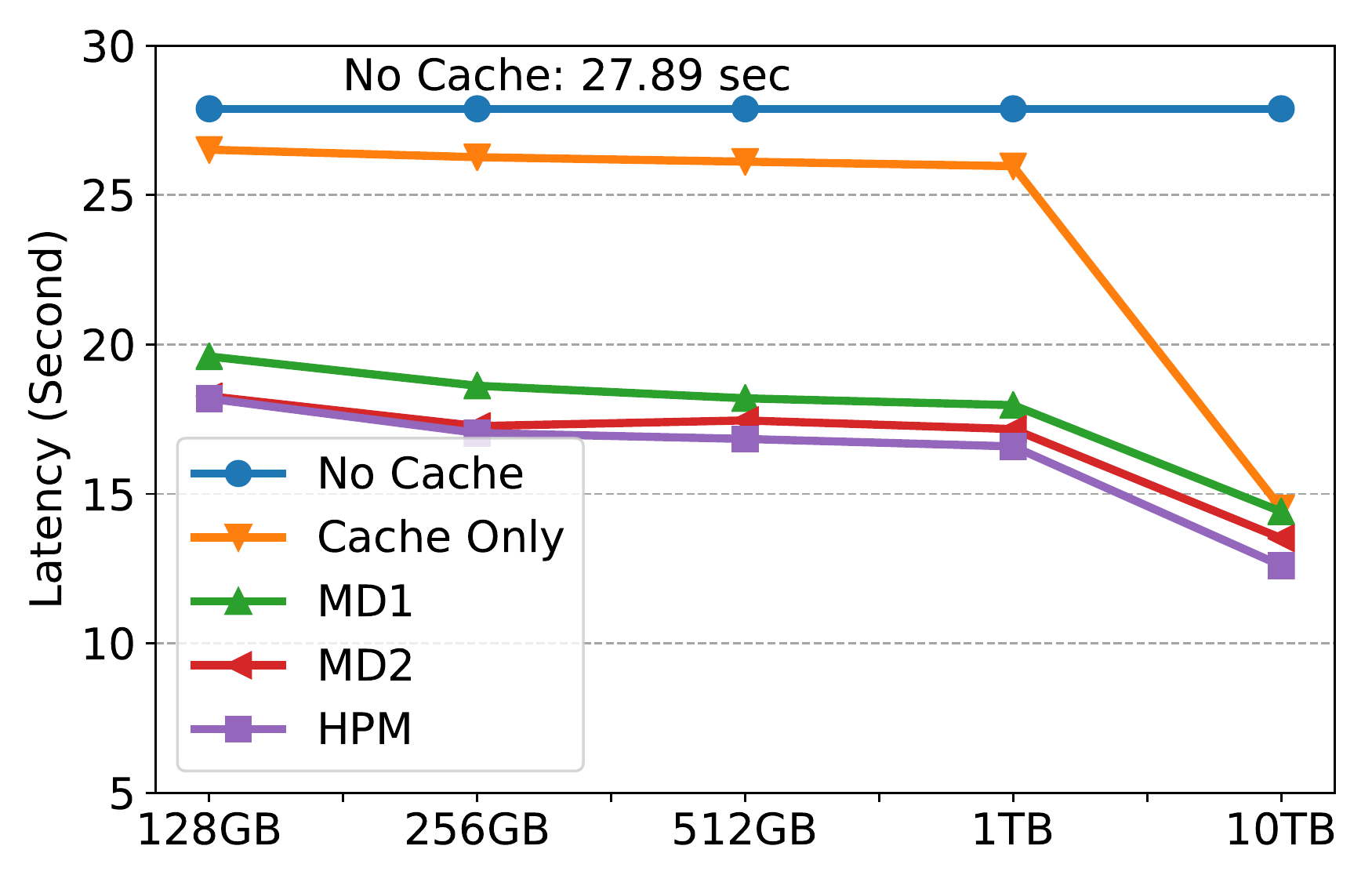}
	\caption{Latency}
	\label{fig:ooi_cache_latency_lru}
\end{subfigure}
\begin{subfigure}{0.3\textwidth}
	\includegraphics[width=\textwidth]{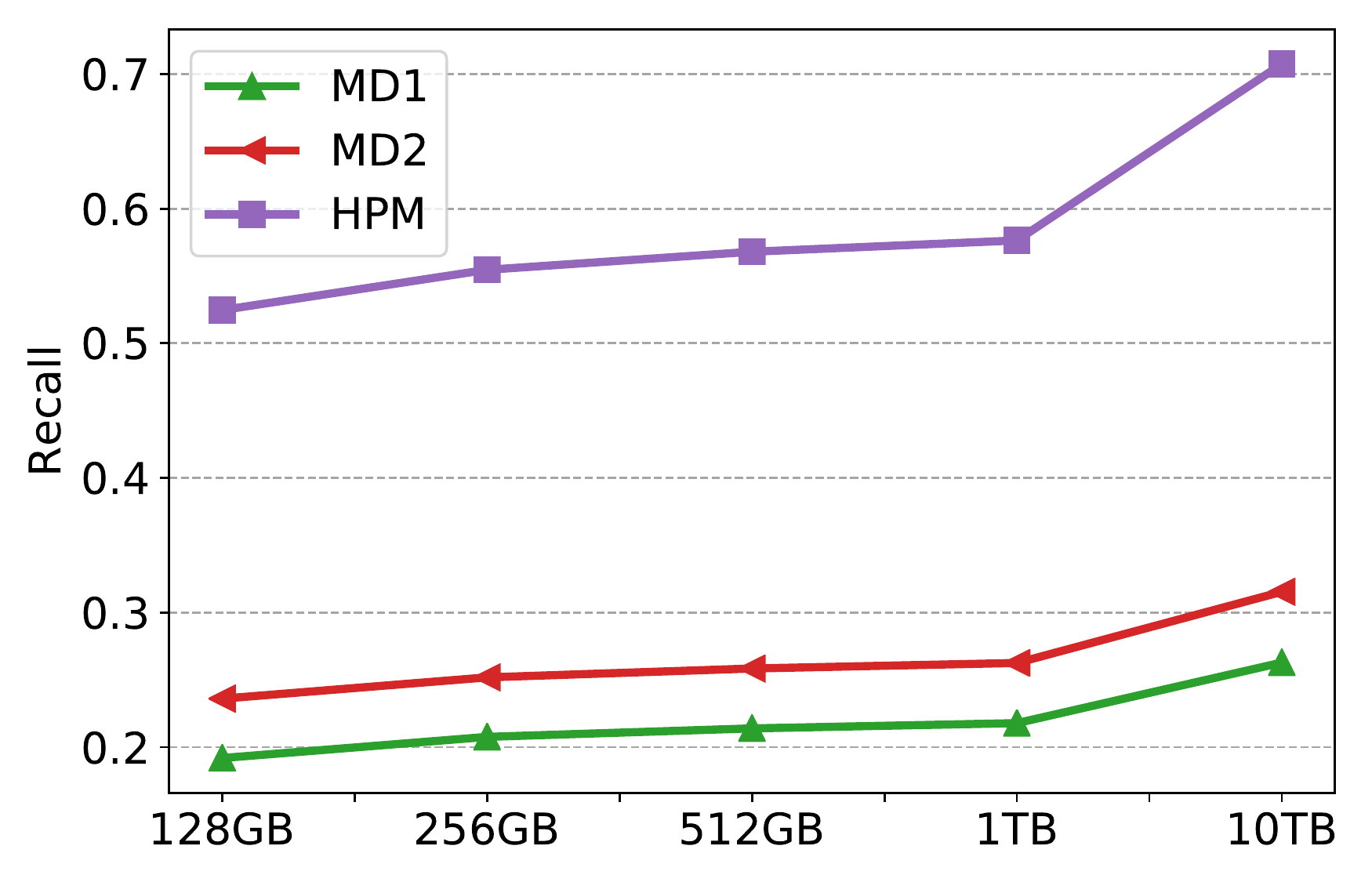}
	\caption{Pre-fetching Recall}
	\label{fig:ooi_recall_lru}
\end{subfigure}	
\caption{OOI LRU cache performance.}
\label{fig:ooi_cache_lru}
\end{figure*}

\begin{figure*}[!h]
\centering
\begin{subfigure}{0.3\textwidth}
	\includegraphics[width=\textwidth]{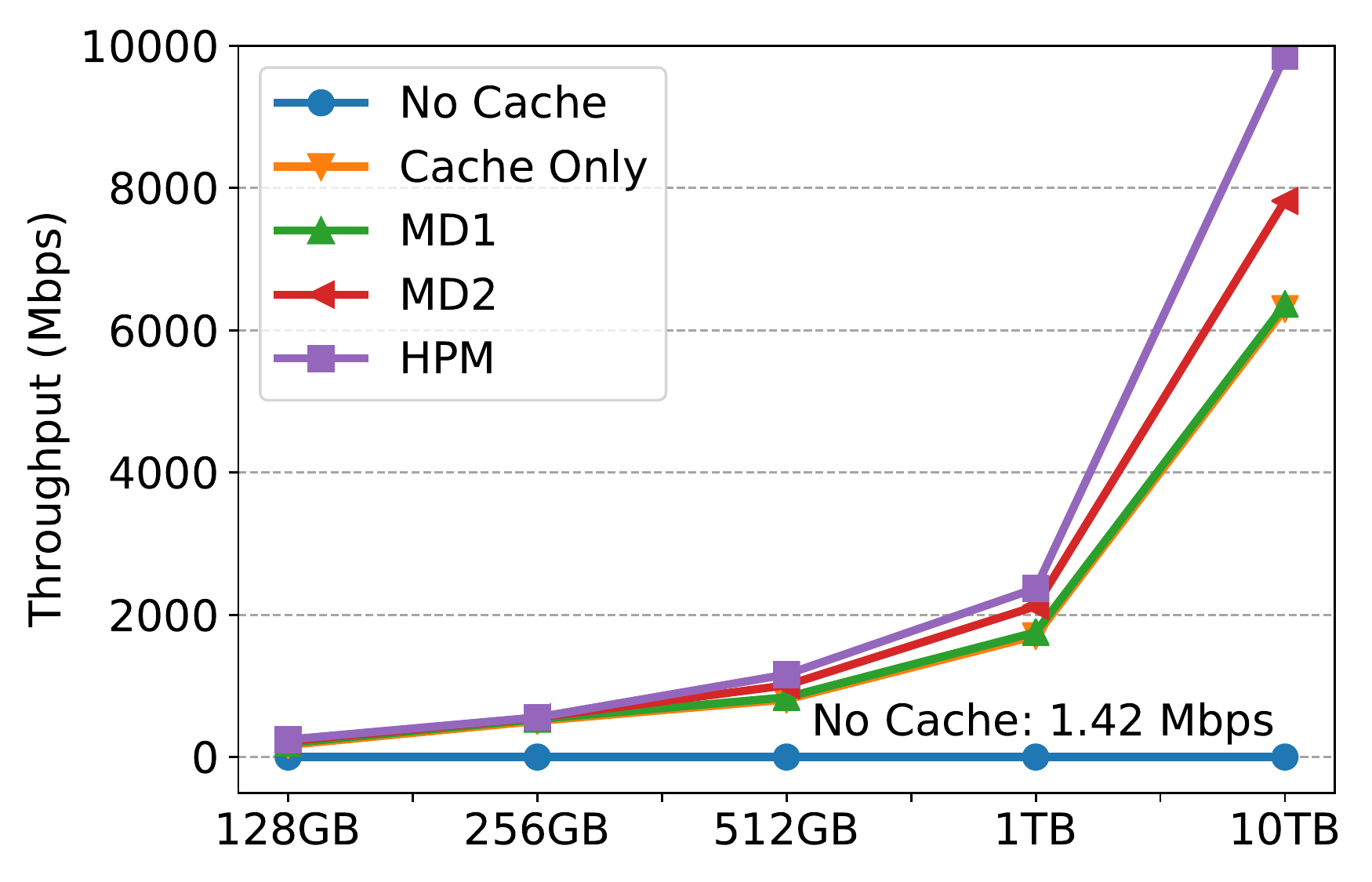}
	\caption{Throughput}
	\label{fig:ooi_cache_throughput_lfu}
\end{subfigure}
\begin{subfigure}{0.3\textwidth}
	\includegraphics[width=\textwidth]{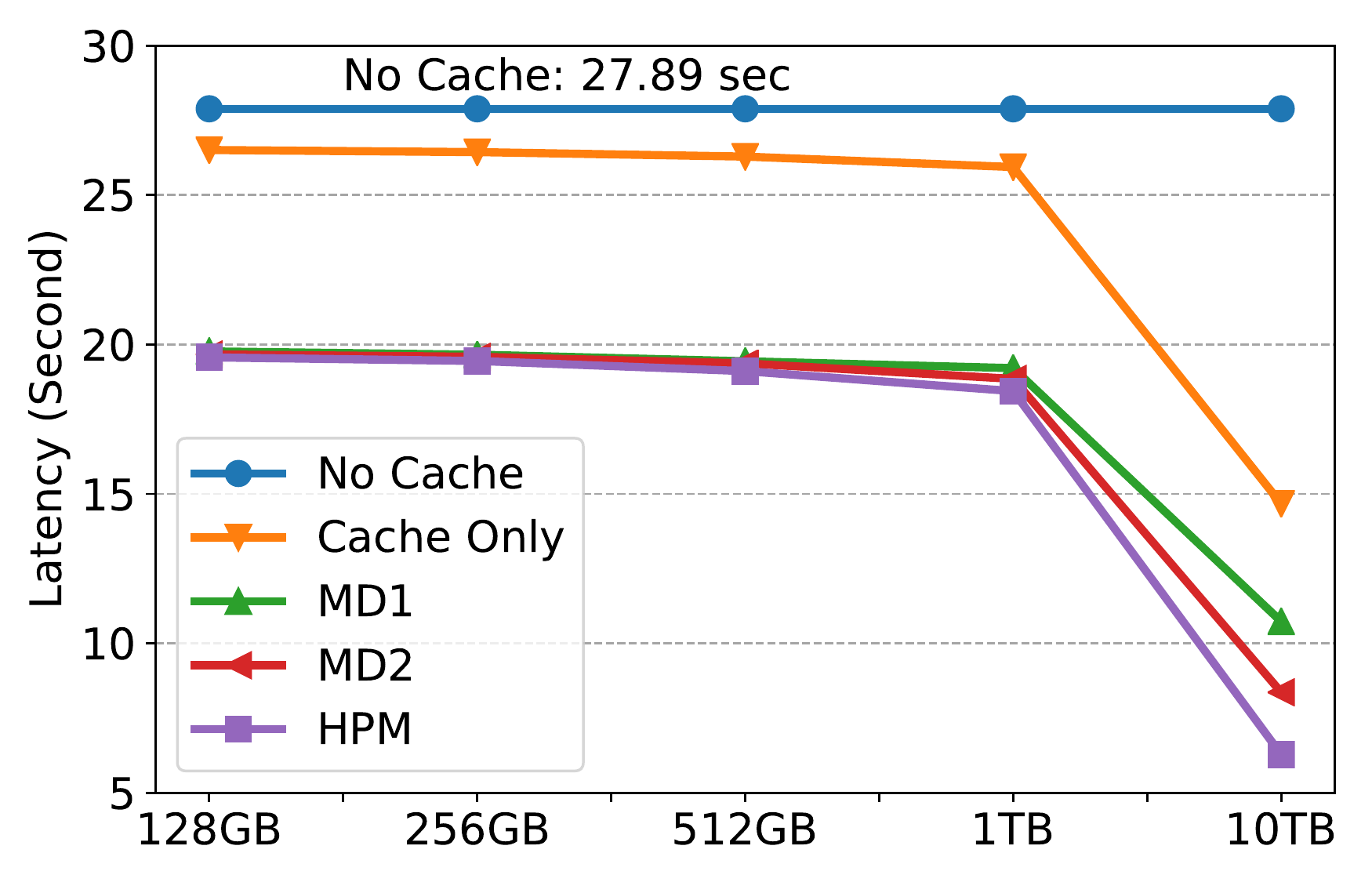}
	\caption{Latency}
	\label{fig:ooi_cache_latency_lfu}
\end{subfigure}
\begin{subfigure}{0.3\textwidth}
	\includegraphics[width=\textwidth]{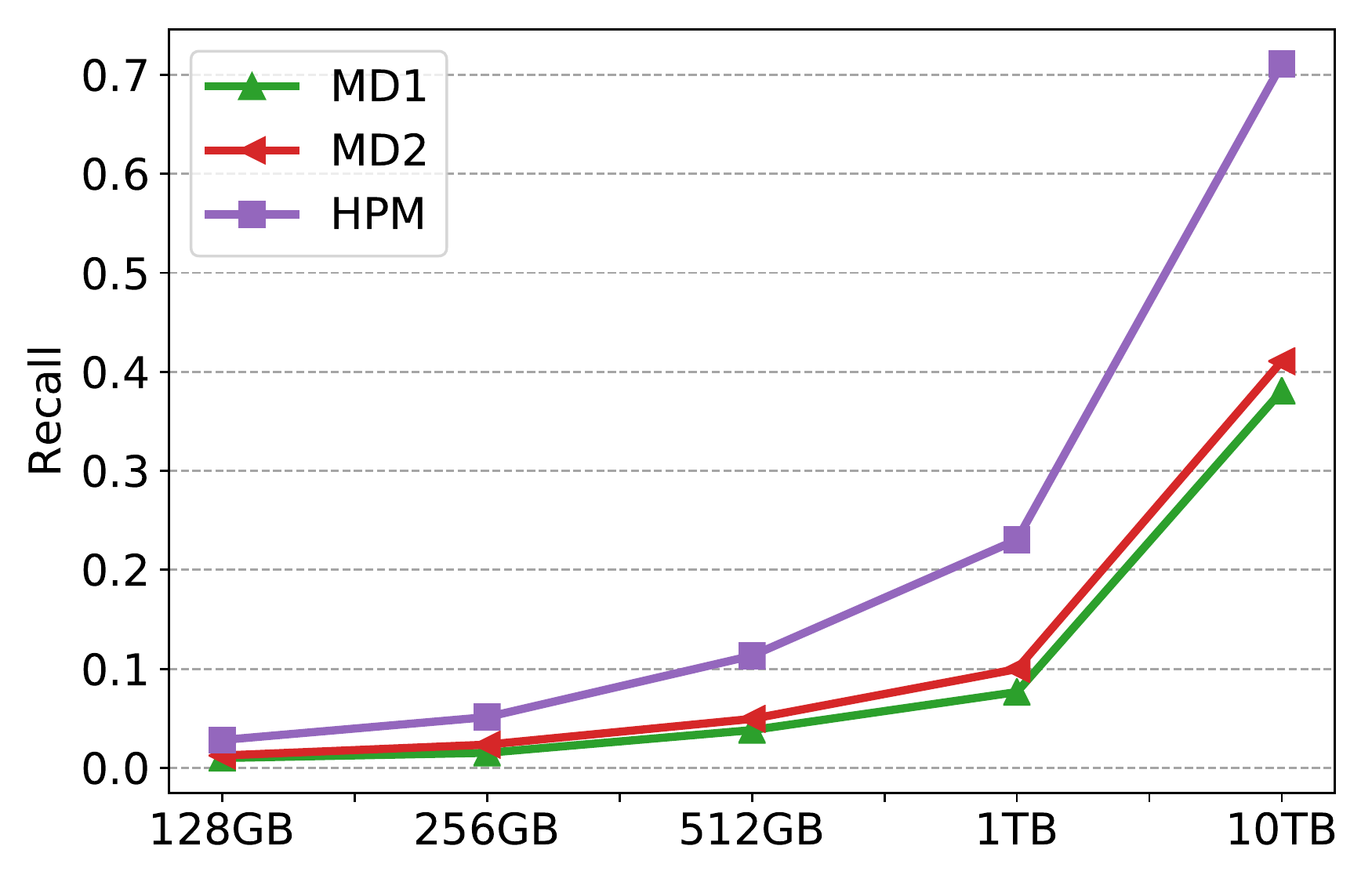}
	\caption{Pre-fetching Recall}
	\label{fig:ooi_recall_lfu}
\end{subfigure}	
\caption{OOI LFU cache performance.}
\label{fig:ooi_cache_lfu}
\end{figure*}

\begin{figure*}[!h]
\centering
\begin{subfigure}{0.3\textwidth}
	\includegraphics[width=\textwidth]{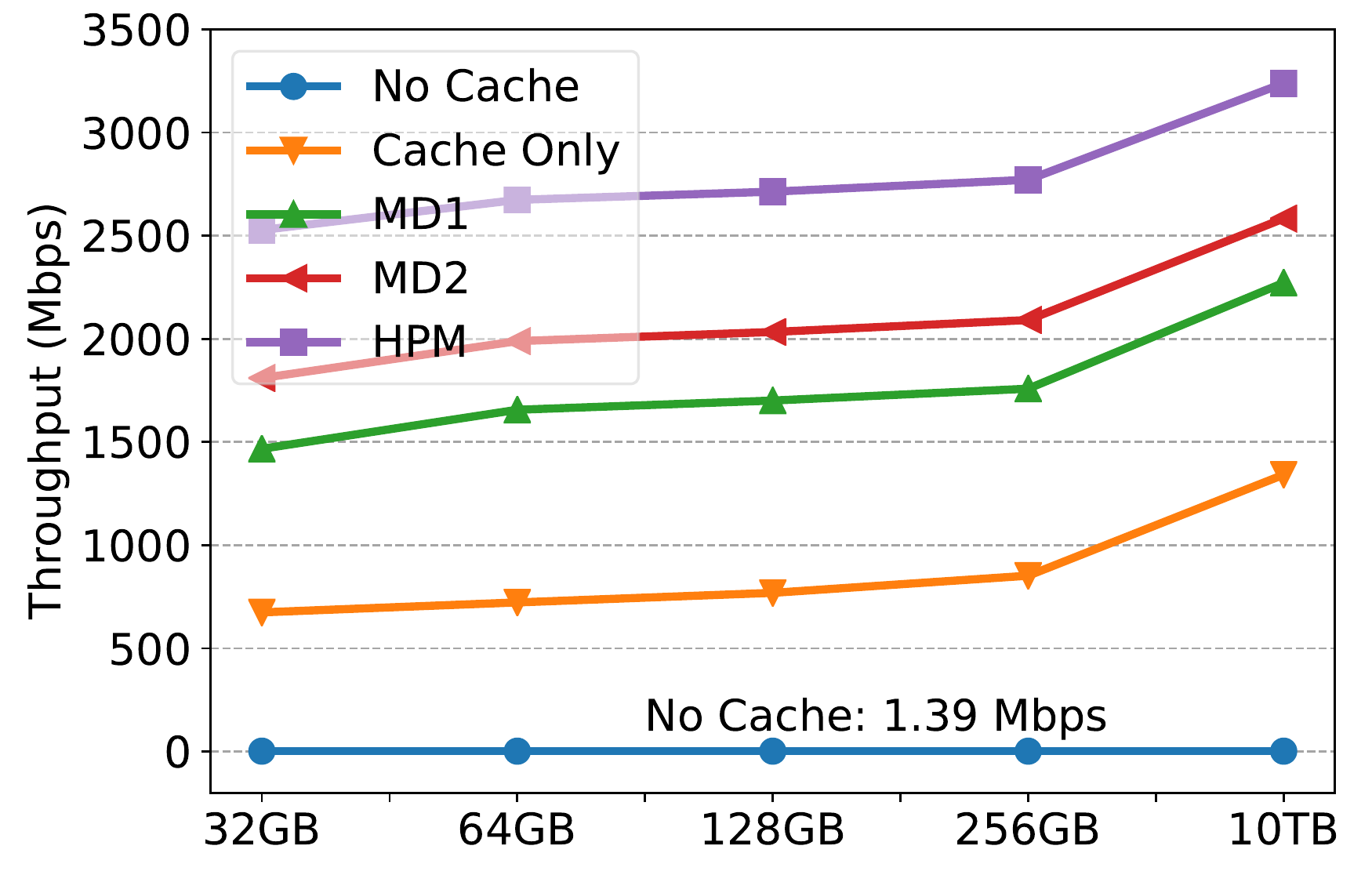}
	\caption{Throughput}
	\label{fig:gage_cache_throughput_lru}
\end{subfigure}
\begin{subfigure}{0.3\textwidth}
	\includegraphics[width=\textwidth]{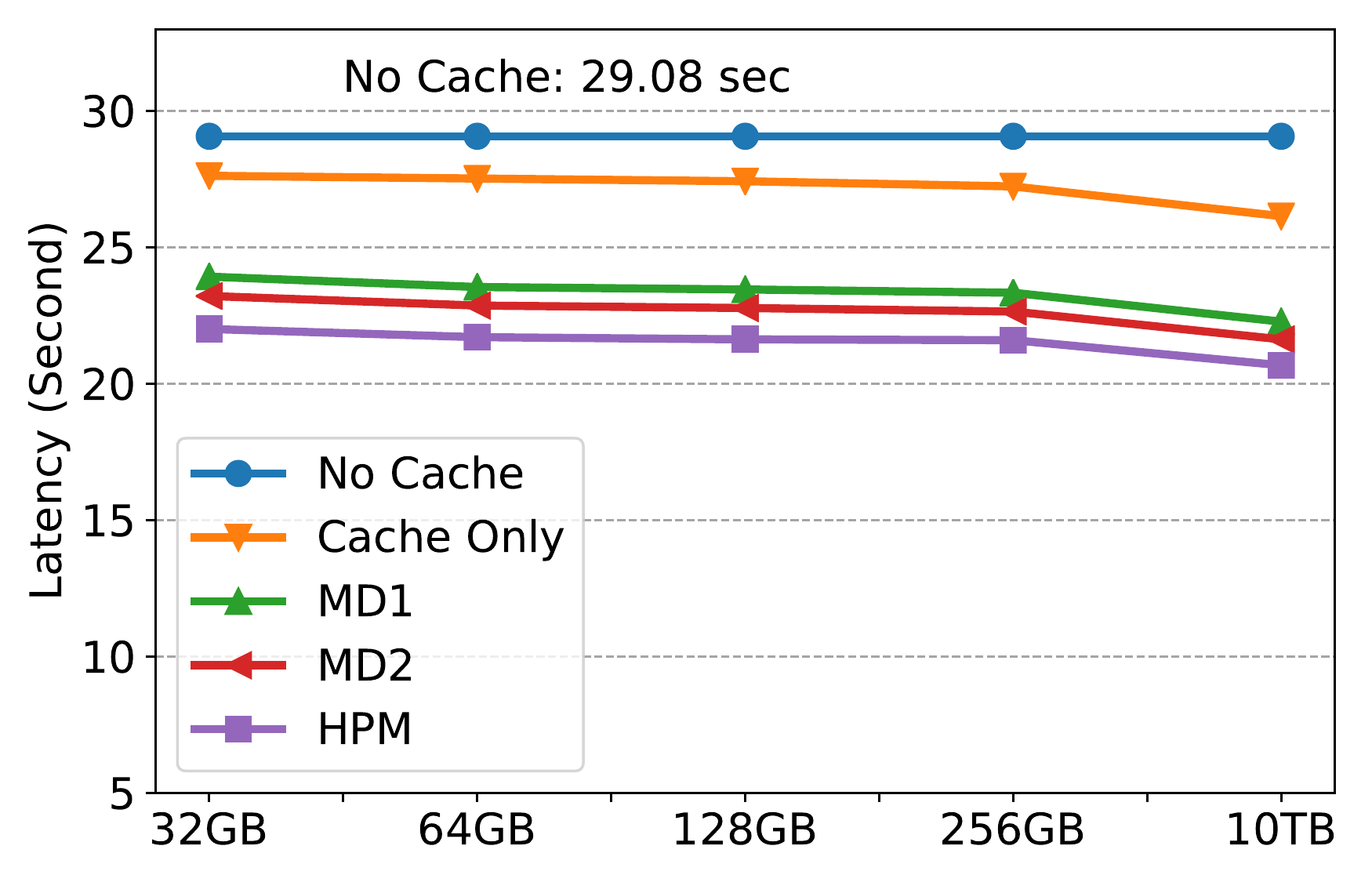}
	\caption{Latency}
	\label{fig:gage_cache_latency_lru}
\end{subfigure}
\begin{subfigure}{0.3\textwidth}
	\includegraphics[width=\textwidth]{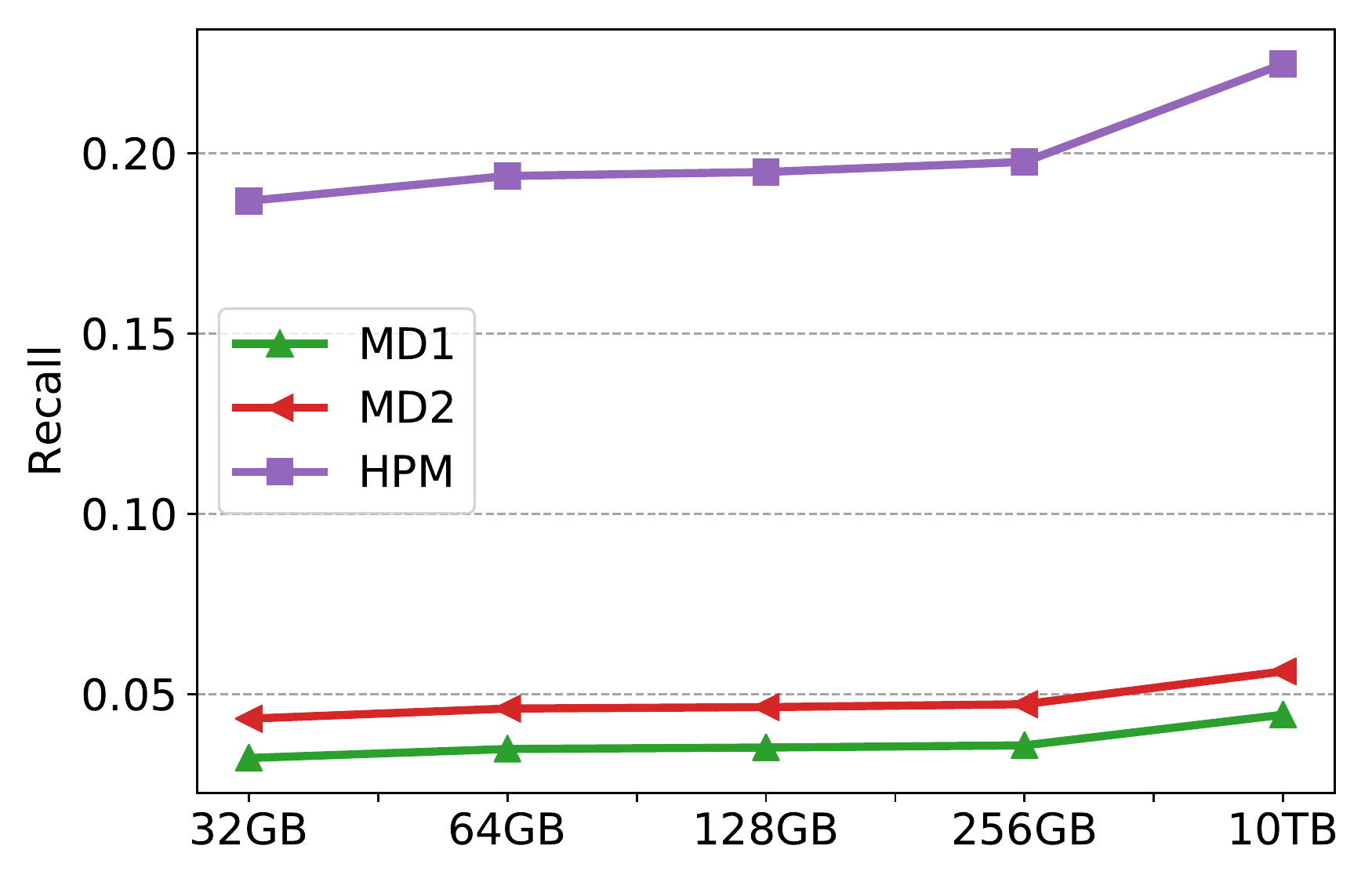}
	\caption{Pre-fetching Recall}
	\label{fig:gage_recall_lru}
\end{subfigure}	
\caption{GAGE LRU cache performance.}
\label{fig:gage_cache_lru}
\end{figure*}

\begin{figure*}[!h]
\centering
\begin{subfigure}{0.3\textwidth}
	\includegraphics[width=\textwidth]{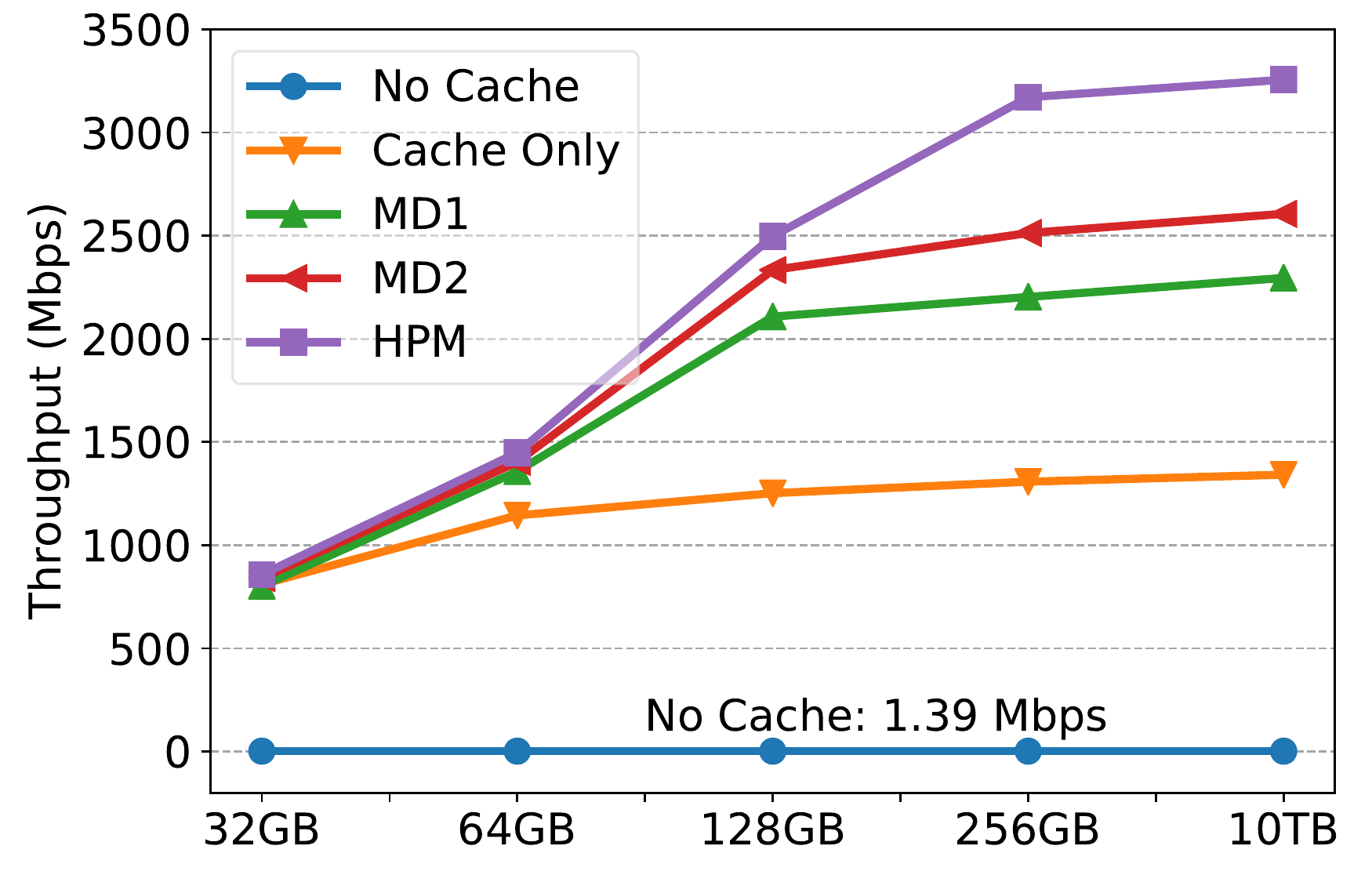}
	\caption{Throughput}
	\label{fig:gage_cache_throughput_lfu}
\end{subfigure}
\begin{subfigure}{0.3\textwidth}
	\includegraphics[width=\textwidth]{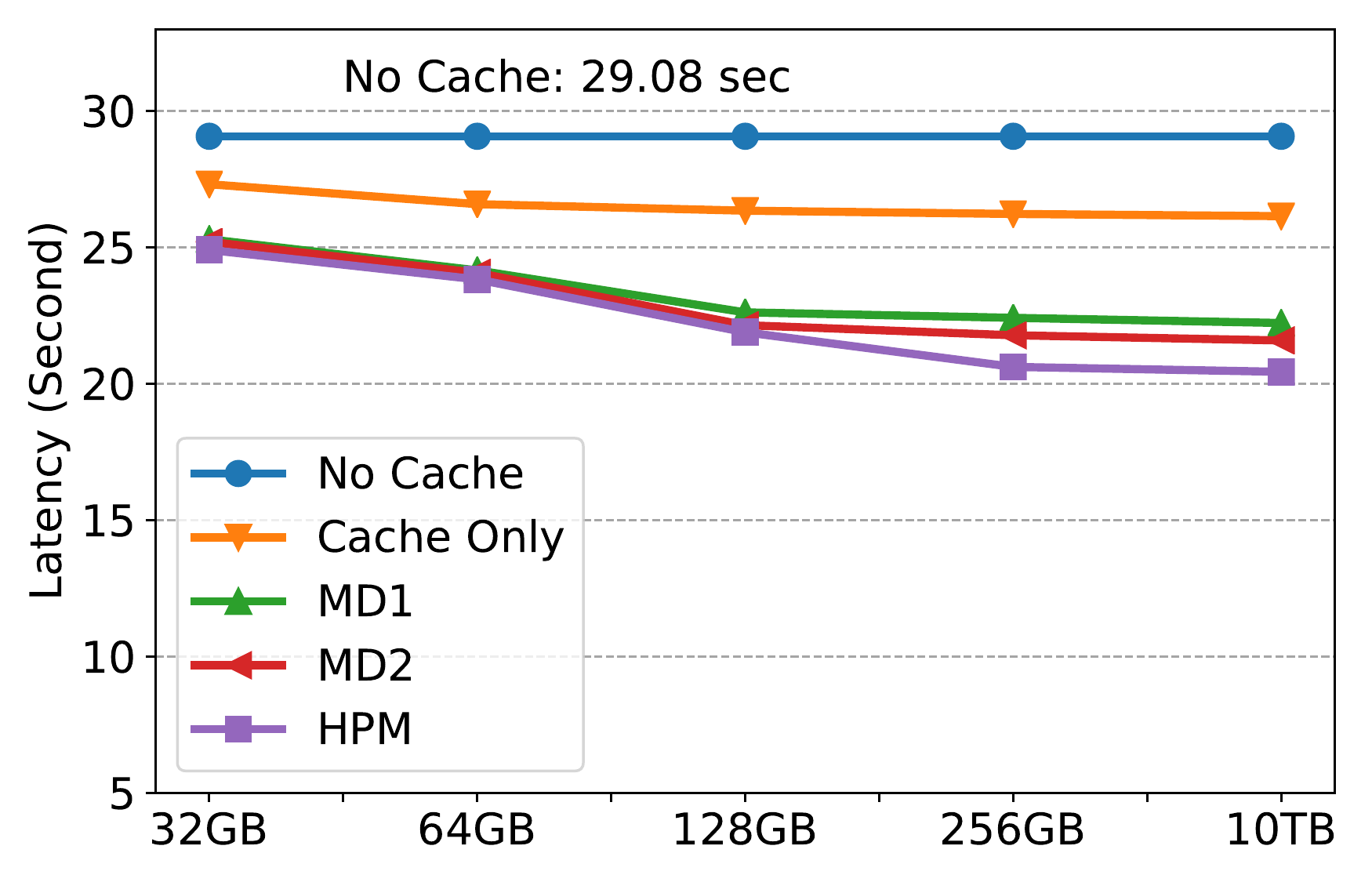}
	\caption{Latency}
	\label{fig:gage_cache_latency_lfu}
\end{subfigure}
\begin{subfigure}{0.3\textwidth}
	\includegraphics[width=\textwidth]{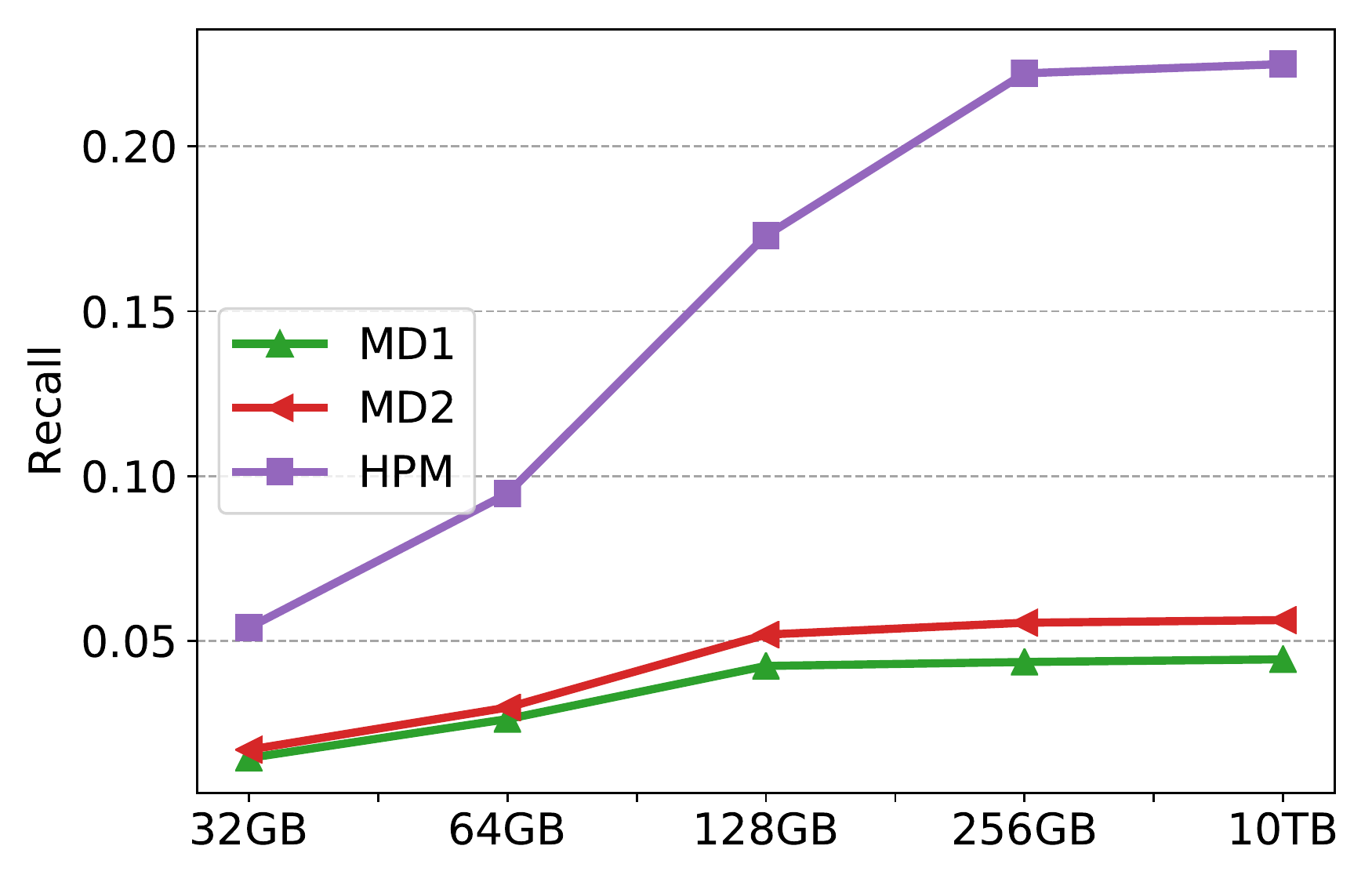}
	\caption{Pre-fetching Recall}
	\label{fig:gage_recall_lfu}
\end{subfigure}	

\caption{GAGE LFU cache performance.}
\label{fig:gage_cache_lfu}
\end{figure*}

\subsection{Experimental results}

\subsubsection{Data delivery performance}

This experiment compares the framework's data delivery performance considering the cache configurations described in Section~\ref{sec:cache_configuration}. Specifically, it compares the \textit{throughput} and \textit{latency} under the following conditions:

\begin{itemize}
    \item $W/O \ Cache$: Represents the current observatory data delivery method that processes all requests by directly transferring data from the observatory to the user.
    \item $Cache \ Only$: Represents the baseline for our framework. It adds a cache layer using the DTNs, but does not use any optimization strategy.
    \item \textit{HPM, MD1, MD2}: Represent our proposed framework that includes a cache layer and data placement optimization strategies, and is configured with three different pre-fetching models.
\end{itemize}

Our framework with pre-fetching achieves significant improvement in accelerating the data delivery performance, where the experiments' results are plotted in Figures~\ref{fig:ooi_cache_lru}-\ref{fig:gage_cache_lfu}. For example, Figure~\ref{fig:ooi_cache_lru} shows that, in case of OOI, using the smallest cache size (128GB), \textit{HPM} increases data transfer \textit{throughput} by 2,689.8x, and reduces request \textit{latency} by 34.8\%  as compared to the $No \ Cache$ case. Only adding a cache layer (i.e.,  the $Cache \ Only$ case), we observe a 739.6x improvement in \textit{throughput}. This confirms our analysis presented in Section~\ref{Sec:observatory_usage_analysis} that indicated significant overlap across request. It also validates the benefits of a cache layer as 
a simple yet effective method for improving the data delivery performance. 

Using pre-fetching results in a 3.6x increase in \textit{throughput} relative to the baseline.  Moreover, \textit{HPM} performs better than the reference models (\textit{MD1}, \textit{MD2}) in all cases. As seen in Figure~\ref{fig:ooi_recall_lru}-\ref{fig:gage_recall_lfu}, the pre-fetching \textit{recall} values indicates that \textit{HPM} achieves the best prediction accuracy. This is because there are over 90.1\% program requests in the traces (Table~\ref{table:user_population_data_volume_analysis}) and the \textit{HPM}'s history-based prediction model can predict these requests with high accuracy. In contrast, the reference models (\textit{MD1} and \textit{MD2}) treat all requests equally, which results in at least 66.7\% wasted pre-fetching and redundant data movement. This validates the benefits of using user data access patterns to improve the data delivery performance.

Furthermore, \textit{MD2} has a higher \textit{recall} value than \textit{MD1} for all cases. This indicates that the association rule-based prediction model performs better than the Markov-based method in predicting  user data request for observatories. It also suggests that \textit{HPM}'s associated rule-based sub-prediction model performs well for non-program requests.

Figures~\ref{fig:ooi_cache_latency_lru}-\ref{fig:gage_cache_latency_lfu} show that our framework decreases \textit{latency} by 29.7\% (OOI, LRU, 128GB) as compared to the $No \ Cache$ case. This is because the cache layer, and the pre-fetching and data streaming mechanisms reduce the number of data requests that are sent to the server DTN. Hence, it reduces the size of the task queue and allows each request to be processed faster. Specifically, Table~\ref{table:num_request} lists the normalized count of requests arriving at the server DTN for each configuration. $HPM$ reduces the number of requests served by the observatory in all cases, especially as compared to the $Cache \ Only$ case. This validates the benefits of using the data push, data pre-fetching  and data streaming mechanisms for \textit{real-time requests}, allowing users to mostly retrieve data from their local DTN.

\begin{table}
\centering
\resizebox{\columnwidth}{!}{%
\begin{tabular}{rrccccc} 
\toprule
\multicolumn{2}{r}{}        & No Cache & Cache Only & MD1    & MD2    & HPM     \\ 
\hline\hline
\\[-0.9em]
\multirow{2}{*}{OOI}  & LRU & 1.0000    & 0.5722      & 0.4653 & 0.4228 & 0.3928  \\
                      & LFU & 1.0000    & 0.9881      & 0.9874 & 0.9871 & 0.9849  \\
\multirow{2}{*}{GAGE} & LRU & 1.0000    & 0.9437      & 0.8707 & 0.8447 & 0.7912  \\
                      & LFU & 1.0000    & 0.9107      & 0.8941 & 0.8905 & 0.8893  \\
\bottomrule
\end{tabular}
}
\caption{Normalized number of user requests for the OOI and GAGE traces that have to be served by the observatory.}
\label{table:num_request}
\end{table}

The results show that the LRU cache eviction policy works better than  the LFU eviction policy for small cache size configurations. LRU achieves 79.6\% (OOI, 128GB) and 91.2\% (GAGE, 32GB) higher \textit{throughput} than LFU using 1TB and 256GB cache sizes respectively. In contrast, the \textit{throughput} curve rises steeply when using LFU  and achieves higher throughput than LRU with large cache sizes (i.e., 10TB, OOI). This implies that the recency of requests is more relevant than the frequency of request for observatories. Since DTN storage resources are limited and considering the typical scale of observatory data, performing well with a small cache size is important, and as a result, we suggest that using the LRU eviction policy is a better choice. We do not further evaluate advanced recency-based eviction models~\cite{ali2012intelligent,ali2014performance} in this paper
and consider it as future work.

\subsubsection{Analysis of the pre-fetching mechanism}

\begin{figure}[t]
\centering
\begin{subfigure}{0.45\textwidth}
	\includegraphics[width=\textwidth]{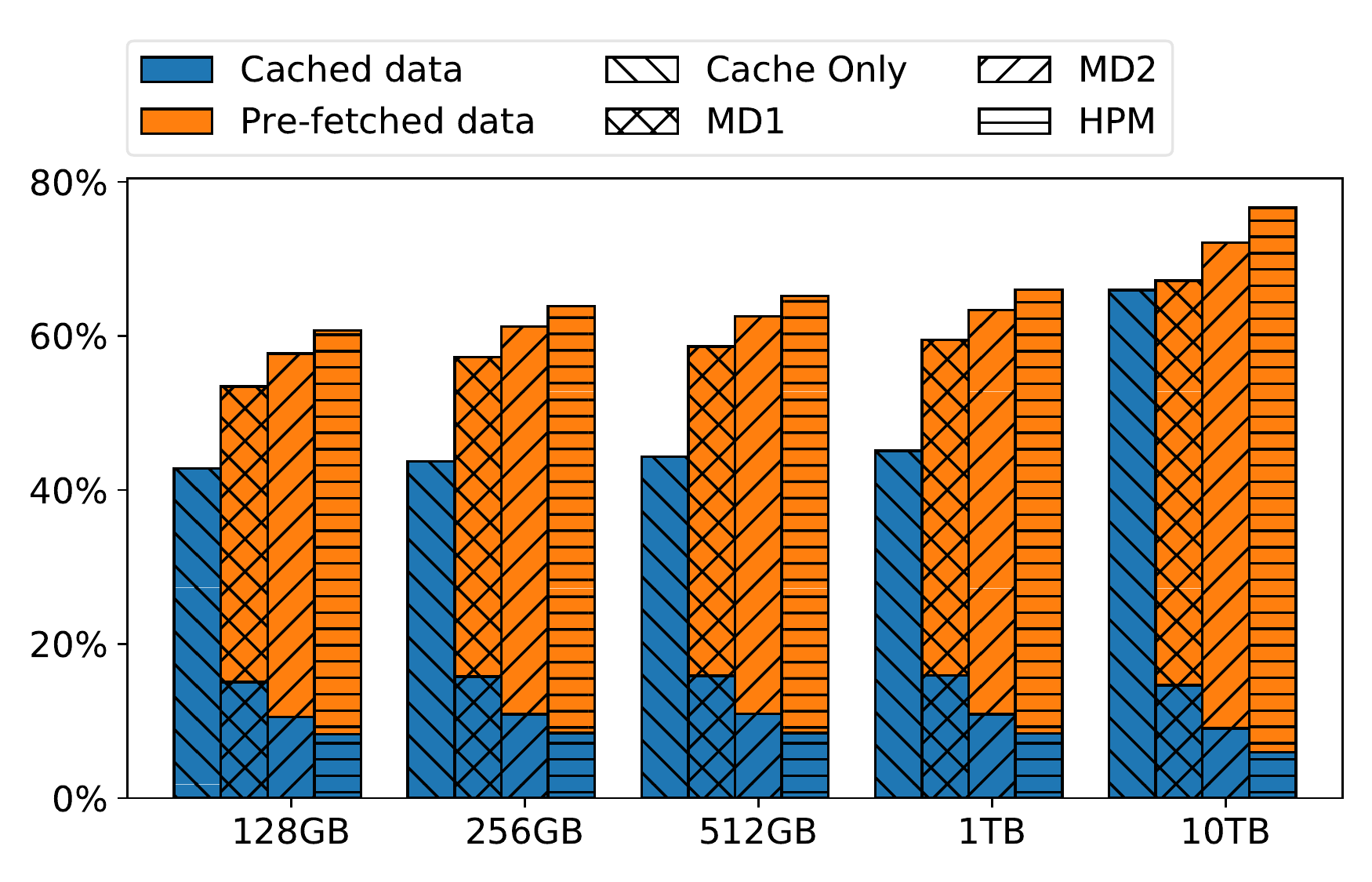}
	\caption{OOI}
	\label{fig:ooi_cache_hit}
\end{subfigure}

\begin{subfigure}{0.45\textwidth}
	\includegraphics[width=\textwidth]{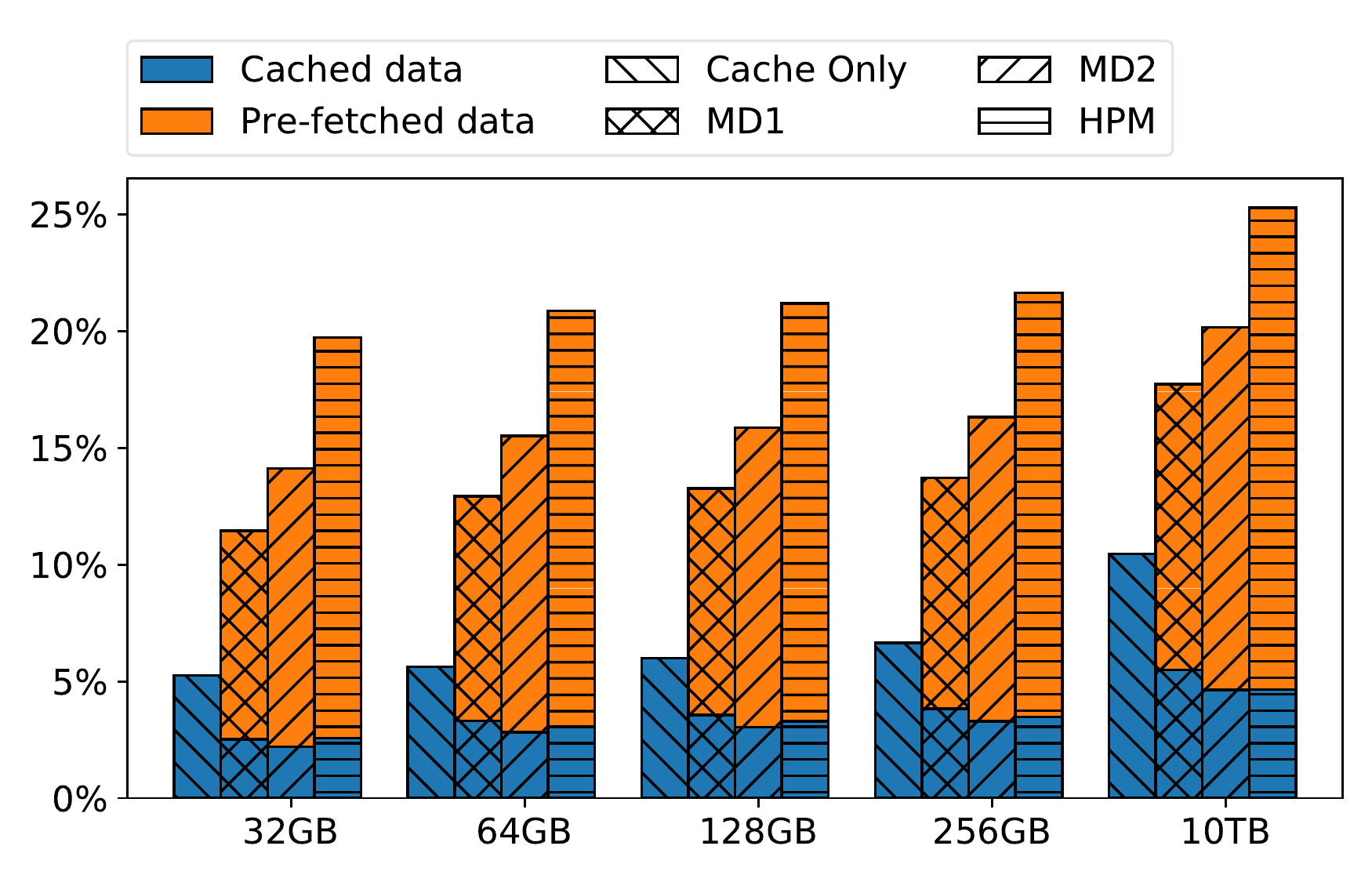}
	\caption{GAGE}
	\label{fig:gage_cache_hit}
\end{subfigure}

	\caption{Data movement from the local cache. These plots shows the percentage of requests that are served by cached and pre-fetched data.}
	\label{fig:data_movement}
\end{figure}

The goal of these experiments is to understand how pre-fetching improves the local data reuse. Figure~\ref{fig:data_movement} plots the average percentage of requests that are served from the local cache for the four strategies using the last experiment for the LRU cache configurations. The color in the plots marks the data sources, where blue is the percentage of requests served using cached data, and the yellow is percentage of requests serviced using pre-fetched data. These plots illustrate that pre-fetching enables users to obtain more data from their local cache. For instance, using the smallest cache sizes (OOI 128GB, GAGE 32GB) and the \textit{HPM} pre-fetching model, the percentage of local data access is 41.9\% and 278.8\% higher than the $Cache \ Only$ case for OOI and GAGE, respectively.

As opposed to passively searching cached data, the pre-fetching mechanism proactively pushes data toward to user. It ensures that users can access more of the their data locally regardless of whether the data is reused from the previous requests. For example, in Figure~\ref{fig:data_movement}b, GAGE has a smaller number of \textit{overlapping requests}; however, the pre-fetching mechanism still enables a much higher percentage of local data accesses as compared to the baseline. 

Furthermore, the pre-fetching mechanism can achieve near-optimal performance with a small cache size. 
The best performance is obtained for a 10TB cache size as the entire datasets for both traces fit into the cache. \textit{HPM} with the smallest cache size can achieve 79.2\% and 78.0\% of this best performance, for OOI and GAGE, respectively.

Based on these experiments, we conclude that our push-based data delivery framework can enable users to access more data from their local DTN cache by proactively pushing data toward users. 

\subsubsection{Evaluation of the data placement strategy}

\begin{table}[!t]
\centering
\resizebox{\columnwidth}{!}{%
\begin{tabular}{rrcccc|l} 
\toprule
\multicolumn{2}{r}{}                          & 32GB     & 64GB     & 128GB    & 256GB    & Avg.      \\ 
\hline\hline
\\[-0.9em]
\% Data opt. by DP                   &        & 21.11\%  & 17.13\%  & 12.95\%  & 10.91\%  & 15.52\%   \\ 
\hline
\\[-0.9em]
\multirow{2}{*}{Throughput (Mbps)}      & W/O DP & 16968.15 & 16899.72 & 16763.17 & 16644.47 & 16818.88  \\
                                     & W/ DP  & 20291.43 & 20230.38 & 20095.91 & 19985.99 & 20150.93  \\
Improv. \%                           &        & 19.59\%  & 19.71\%  & 19.88\%  & 20.08\%  & 19.81\%   \\ 
\hline
\\[-0.9em]
\multirow{2}{*}{Tot. perf. improv. (Mbps)} & W/O DP & 2446.10  & 2603.19  & 2658.92  & 2723.77  & 2608.00   \\
                                     & W/ DP  & 2527.70  & 2673.68  & 2713.45  & 2771.17  & 2671.50   \\
Tot. perf. improv. \%                      &        & 3.34\%   & 2.71\%   & 2.05\%   & 1.74\%   & 2.46\%    \\
\bottomrule
\end{tabular}
}
\caption{Impact of the data placement strategy (DP): Percentage of cached data optimized by the DP, throughput (Mbps) for retrieving the data from a peer DTN's cache, and the impact of the DP on total data transfer performance.}
\label{table:data_placement}
\end{table}

In this experiment, we use \textit{HPM} with LRU and the GAGE trace to evaluate the impact of the data placement strategy on the data delivery performance. 
Table~\ref{table:data_placement} summarizes the results. The first row presents the percentage of cached data that has been optimized by the data placement strategy using different cache configurations. We observe that the data placement strategy is more effective for a small cache size because it replicates hot data to the local data hub.

We also measure the average throughput for retrieving data from the peer DTN cache. The data placement strategy improves throughput by 19.81\% on average. The performance increases slightly as the cache size gets larger because the local data hub can keep the data replica longer. Overall, the data placement strategy improves data delivery performance by 2.46\% on average.
We anticipate that our data placement strategy is likely to impact overall data delivery performance even more significantly as the scale of the DTN network grows, the network gets more heterogeneous and complex, and more the amount data retrieved from peer DTN caches increases.

\subsubsection{Framework performance for different network condition and request traffic}

\begin{table*}[!t]
\centering
\begin{subtable}[t]{\textwidth}
\centering
\resizebox{\textwidth}{!}{%
\begin{tabular}{rccccclccccclccccc} 
\toprule
\multirow{2}{*}{Network} & \multicolumn{5}{c}{Low Request Traffic}               &  & \multicolumn{5}{c}{Regular Request Traffic}           &  & \multicolumn{5}{c}{Heavy Request Traffic}              \\ 
\cline{2-6}\cline{8-12}\cline{14-18}
\\[-0.9em]
                         & No Cache & Cache Only & MD1     & MD2     & HPM     &  & No Cache & Cache Only & MD1     & MD2     & HPM     &  & No Cache & Cache Only & MD1     & MD2     & HPM      \\ 
\hline\hline
\\[-0.9em]
Best                     & 2.85      & 1326.97     & 3849.01 & 4881.04 & 5609.05 &  & 1.42      & 1322.24     & 3509.73 & 4221.31 & 4705.25 &  & 1.21      & 1311.97     & 2680.96 & 3201.53 & 3582.40  \\
Medium                    & 2.01      & 1327.27834  & 3858.24 & 4899.32 & 5615.12 &  & 1.11      & 1319.97     & 3506.01 & 4221.02 & 4707.16 &  & 0.92      & 1312.27     & 2677.13 & 3200.63 & 3582.07  \\
Worst                    & 1.01      & 1326.95858  & 2507.73 & 3185.79 & 3648.89 &  & 0.71      & 1318.08     & 2278.10 & 2743.75 & 3057.84 &  & 0.58      & 1311.96     & 1739.93 & 2080.23 & 2327.72  \\
\bottomrule
\end{tabular}
}
\caption{OOI throughput for different network conditions and request traffic levels using LRU.}
\label{tab:ooi_network_comparison_throughput}
\end{subtable}

\hspace{\fill}

\begin{subtable}[t]{\textwidth}
\centering
\resizebox{\textwidth}{!}{%
\begin{tabular}{rccccclccccclccccc} 
\toprule
\multirow{2}{*}{Network} & \multicolumn{5}{c}{Low Request Traffic}               &  & \multicolumn{5}{c}{Regular Request Traffic}           &  & \multicolumn{5}{c}{Heavy Request Traffic}              \\ 
\cline{2-6}\cline{8-12}\cline{14-18}
\\[-0.9em]
                         & No Cache & Cache Only & MD1     & MD2     & HPM     &  & No Cache & Cache Only & MD1     & MD2     & HPM     &  & No Cache & Cache Only & MD1     & MD2     & HPM      \\ 
\hline\hline
\\[-0.9em]
Best                     & 1.52      & 851.84   & 1758.36 & 2091.16 & 2771.17 &  & 1.39      & 850.00      & 1501.11 & 1903.03 & 2488.08 &  & 1.31      & 849.02      & 1147.16 & 1348.49 & 1729.56  \\
Medium                    & 1.26      & 853.12   & 1760.41 & 2092.90 & 2771.26 &  & 1.19      & 851.99      & 1498.31 & 1904.62 & 2487.06 &  & 1.09      & 848.52      & 1146.77 & 1351.36 & 1730.24  \\
Worst                    & 0.69      & 853.45   & 1214.21 & 1443.12 & 1908.88 &  & 0.69      & 849.97      & 1035.21 & 1313.22 & 1715.93 &  & 0.60      & 849.50      & 789.01  & 927.91  & 1193.97  \\
\bottomrule
\end{tabular}
}

\caption{GAGE throughput for different network conditions and request traffic levels using LRU.}
\label{tab:gage_network_comparison_throughput}
\end{subtable}

\caption{Pre-fetching model performance comparison for different network conditions and request traffic levels. HPM is our hybrid pre-fetching model; DM1 and DM2 are the reference pre-fetching  models.}
\label{tab:network_comparison_throughput}
\end{table*}

The goal of this experiment is to evaluate the ability of the push-based data delivery framework presented in this paper to tolerate different network conditions and request traffic levels. 
As presented in Section~\ref{sec:variations}, we create three network conditions, i.e., \textit{best}, \textit{medium} and \textit{worst}, where \textit{medium} and \textit{worst} are 50\% and 1\% of the \textit{best} case respective, and three levels of request traffic. Moreover, we use a LRU cache of size 1TB and 256GB for the OOI and GAGE traces respectively. We compare the achieved \textit{throughput}, which is presented in Table~\ref{tab:network_comparison_throughput}. 

The columns of the table compares performance of each strategy for different network conditions. The performance remains constant for \textit{HPM}, \textit{MD1} and  \textit{MD2} for the \textit{best} and \textit{medium} network conditions. However, for the \textit{worst} network condition, the throughput drops down by 34.9\% and 31.1\% relative to the \textit{best} case for OOI and GAGE, respectively. 
These results illustrate that the pre-fetching mechanism tolerates network bandwidth variations as it can exploit the potentially idle network resource to transfer data in advance. We can also see that the network conditions significantly impact the $No \ Cache$ case as in this case data needs to downloaded directly from the observatory. This  indicates that the data delivery methods currently implemented by observatories are not very resilient to unfavorable network conditions. Conversely, the $Cache \ Only$ strategy is not significantly affected by the network conditions because in this case, data is mostly retrieved from the local DTN's cache.

The rows of the table present the average data transfer throughput for each cache strategy and for the \textit{low} to \textit{high} request traffic levels. The request traffic impacts all the cases except $Cache \ Only$ as there is limited concurrent processing in this case. Since our simulations use a fixed number (\textit{ten}) of service processes, heavier request traffic implies longer queuing time. Moreover, given sufficient processing capability, multiple concurrent data transfers decrease the shared bandwidth available to each task. 
As a result, the data transfer time increases, which can result in the pre-fetched data arriving too late and being wasted. 
We conclude that controlling the request traffic at the observatory is important. The push-based data delivery framework presented in this paper addresses this by using  caching and pre-fetching mechanisms to reduce the request traffic.

\section{Conclusion}
\label{Sec:conclusion}
The paper addressed data access challenges for large-scale, share-use facilities such as instruments, observatories and experimental platforms with the goal of improving data access performance and overall user experience for these facilities. Specifically, in this work we analyze the data access and usage patterns for two large-scale facilities, OOI and GAGE and study how these patterns can be leveraged to predict future requests.

In this paper, we have presented a push-based data delivery framework that leverages user access patterns to design a hybrid data pre-fetching model along with optimized cache mechanisms aimed at accelerating data delivery performance for distributed observatories. Furthermore, we presented an implementation of this framework that leverages the VDC Science DMZ and uses in-network DTNs to host the data pre-fetching and caching services. 

We evaluated our framework using a simulated VDC environment and OOI and GAGE user access logs. The experimental evaluation indicated a significant reduction in the amount of data transferred over the network for both, the OOI and GAGE observatories as  compared to the current approach (60.7\% and 19.7\% improvements for OOI and GAGE respectively). 

Potential future optimizations include replacing the ARIMA time-series prediction model with the portable Recurrent Neural Network (RNN) based predictor~\cite{tseng2019towards}, which could further improve performance. Our current efforts include building a prototype implementation to be deployed in the Virtual Data Collaboratory.

\section*{Acknowledgments}
This research is supported in part by NSF via grants numbers OCE 1745246, OAC 1835692, OAC 1826997, and OAC 1640834, and was conducted as part of the Rutgers Discovery Informatics Institute (RDI$^2$). The materials used are based in part on services provided by the GAGE Facility, operated by UNAVCO, Inc., with support from NSF and the National Aeronautics and Space Administration under NSF Cooperative Agreement EAR-1724794 and NSF grant OAC 1835791. We thank the reviewers for their careful review, which helped improve the manuscript.

\bibliographystyle{IEEEtran}
\bibliography{references}

\begin{thebibliography}{10}
\providecommand{\url}[1]{#1}
\csname url@samestyle\endcsname
\providecommand{\newblock}{\relax}
\providecommand{\bibinfo}[2]{#2}
\providecommand{\BIBentrySTDinterwordspacing}{\spaceskip=0pt\relax}
\providecommand{\BIBentryALTinterwordstretchfactor}{4}
\providecommand{\BIBentryALTinterwordspacing}{\spaceskip=\fontdimen2\font plus
\BIBentryALTinterwordstretchfactor\fontdimen3\font minus
  \fontdimen4\font\relax}
\providecommand{\BIBforeignlanguage}[2]{{%
\expandafter\ifx\csname l@#1\endcsname\relax
\typeout{** WARNING: IEEEtran.bst: No hyphenation pattern has been}%
\typeout{** loaded for the language `#1'. Using the pattern for}%
\typeout{** the default language instead.}%
\else
\language=\csname l@#1\endcsname
\fi
#2}}
\providecommand{\BIBdecl}{\relax}
\BIBdecl

\bibitem{abbott2016observation}
B.~P. Abbott, R.~Abbott, T.~Abbott, M.~Abernathy, F.~Acernese, K.~Ackley,
  C.~Adams, T.~Adams, P.~Addesso, R.~Adhikari \emph{et~al.}, ``Observation of
  gravitational waves from a binary black hole merger,'' \emph{Physical review
  letters}, vol. 116, no.~6, p. 061102, 2016.

\bibitem{akiyama2019first}
K.~Akiyama, A.~Alberdi, W.~Alef, K.~Asada, R.~Azulay, A.-K. Baczko, D.~Ball,
  M.~Balokovi{\'c}, J.~Barrett, D.~Bintley \emph{et~al.}, ``First m87 event
  horizon telescope results. iv. imaging the central supermassive black hole,''
  \emph{The Astrophysical Journal Letters}, vol. 875, no.~1, p.~L4, 2019.

\bibitem{rodero2019data}
I.~Rodero and M.~Parashar, ``Data cyberinfrastructure for end-to-end science,''
  \emph{Computing in Science and Engineering}, vol.~22, no.~05, pp. 60--71,
  2020.

\bibitem{deelman2019cyberinfrastructure}
E.~Deelman, A.~Mandal, V.~Pascucci, S.~Sons, J.~Wyngaard, C.~Vardeman,
  S.~Petruzza, I.~Baldin, L.~Christopherson, R.~Mitchell \emph{et~al.},
  ``Cyberinfrastructure center of excellence pilot: Connecting large facilities
  cyberinfrastructure,'' in \emph{2019 15th International Conference on
  eScience (eScience)}.\hskip 1em plus 0.5em minus 0.4em\relax IEEE, 2019, pp.
  449--457.

\bibitem{aaai2020}
K.~Fauvel, D.~Balouek-Thomert, D.~Melgar, P.~Silva, A.~Simonet, G.~Antoniu,
  A.~Costan, V.~Masson, M.~Parashar, I.~Rodero \emph{et~al.}, ``A distributed
  multi-sensor machine learning approach to earthquake early warning,'' in
  \emph{Proceedings of the AAAI Conference on Artificial Intelligence},
  vol.~34, no.~01, 2020, pp. 403--411.

\bibitem{dewdney2009square}
P.~Dewdney, P.~Hall, R.~Schillizzi, and J.~Lazio, ``The square kilometre
  array,'' \emph{Proceedings of the Institute of Electrical and Electronics
  Engineers IEEE}, vol.~97, no.~8, pp. 1482--1496, 2009.

\bibitem{abramovici1992ligo}
A.~Abramovici, W.~E. Althouse, R.~W. Drever, Y.~G{\"u}rsel, S.~Kawamura, F.~J.
  Raab, D.~Shoemaker, L.~Sievers, R.~E. Spero, K.~S. Thorne \emph{et~al.},
  ``Ligo: The laser interferometer gravitational-wave observatory,''
  \emph{science}, vol. 256, no. 5055, pp. 325--333, 1992.

\bibitem{kampe2010neon}
T.~U. Kampe, B.~R. Johnson, M.~A. Kuester, and M.~Keller, ``Neon: the first
  continental-scale ecological observatory with airborne remote sensing of
  vegetation canopy biochemistry and structure,'' \emph{Journal of Applied
  Remote Sensing}, vol.~4, no.~1, p. 043510, 2010.

\bibitem{deelman2015pegasus}
E.~Deelman, K.~Vahi, G.~Juve, M.~Rynge, S.~Callaghan, P.~J. Maechling,
  R.~Mayani, W.~Chen, R.~F. Da~Silva, M.~Livny \emph{et~al.}, ``Pegasus, a
  workflow management system for science automation,'' \emph{Future Generation
  Computer Systems}, vol.~46, pp. 17--35, 2015.

\bibitem{albrecht2012makeflow}
M.~Albrecht, P.~Donnelly, P.~Bui, and D.~Thain, ``Makeflow: A portable
  abstraction for data intensive computing on clusters, clouds, and grids,'' in
  \emph{Proceedings of the 1st ACM SIGMOD Workshop on Scalable Workflow
  Execution Engines and Technologies}, 2012, pp. 1--13.

\bibitem{roderoenabling}
I.~Rodero, Y.~Qin, J.~Valls, A.~Simonet, J.~Villalobos, M.~Parashar, C.~Youn,
  C.~Wang, K.~Thareja, P.~Ruth \emph{et~al.}, ``Enabling data streaming-based
  science gateways through federated cyberinfrastructure,'' \emph{Gateways
  2019}, 2019.

\bibitem{zamani2017deadline}
A.~R. Zamani, M.~Zou, J.~Diaz-Montes, I.~Petri, O.~Rana, A.~Anjum, and
  M.~Parashar, ``Deadline constrained video analysis via in-transit
  computational environments,'' \emph{IEEE Transactions on Services Computing},
  2017.

\bibitem{renart2019edge}
E.~G. Renart, D.~Balouek-Thomert, and M.~Parashar, ``An edge-based framework
  for enabling data-driven pipelines for iot systems,'' in \emph{2019 IEEE
  International Parallel and Distributed Processing Symposium Workshops
  (IPDPSW)}.\hskip 1em plus 0.5em minus 0.4em\relax IEEE, 2019, pp. 885--894.

\bibitem{balouek2019towards}
D.~Balouek-Thomert, E.~G. Renart, A.~R. Zamani, A.~Simonet, and M.~Parashar,
  ``Towards a computing continuum: Enabling edge-to-cloud integration for
  data-driven workflows,'' \emph{The International Journal of High Performance
  Computing Applications}, vol.~33, no.~6, pp. 1159--1174, 2019.

\bibitem{ali2011survey}
W.~Ali, S.~M. Shamsuddin, A.~S. Ismail \emph{et~al.}, ``A survey of web caching
  and prefetching,'' \emph{Int. J. Advance. Soft Comput. Appl}, vol.~3, no.~1,
  pp. 18--44, 2011.

\bibitem{zhang2013caching}
G.~Zhang, Y.~Li, and T.~Lin, ``Caching in information centric networking: A
  survey,'' \emph{Computer Networks}, vol.~57, no.~16, pp. 3128--3141, 2013.

\bibitem{gagliardi2011content}
J.~D. Gagliardi and T.~S. Munger, ``Content delivery network,'' Jun.~14 2011,
  uS Patent 7,962,580.

\bibitem{jiang2018cachalot}
F.~Jiang, C.~Castillo, and S.~Ahalt, ``Cachalot: A network-aware, cooperative
  cache network for geo-distributed, data-intensive applications,'' in
  \emph{NOMS 2018-2018 IEEE/IFIP Network Operations and Management
  Symposium}.\hskip 1em plus 0.5em minus 0.4em\relax IEEE, 2018, pp. 1--9.

\bibitem{PRP}
L.~Smarr, C.~Crittenden, T.~DeFanti, J.~Graham, D.~Mishin, R.~Moore,
  P.~Papadopoulos, and F.~W\"{u}rthwein, ``The pacific research platform:
  Making high-speed networking a reality for the scientist,'' in
  \emph{Proceedings of the Practice and Experience on Advanced Research
  Computing}, ser. PEARC '18, 2018, pp. 29:1--29:8.

\bibitem{parashar2019virtual}
M.~Parashar, V.~Honavar, A.~Simonet, I.~Rodero, F.~Ghahramani, G.~Agnew, and
  R.~Jantz, ``The virtual data collaboratory,'' \emph{Computing in Science and
  Engineering}, 2019.

\bibitem{altintas2019workflow}
I.~Altintas, K.~Marcus, I.~Nealey, S.~L. Sellars, J.~Graham, D.~Mishin,
  J.~Polizzi, D.~Crawl, T.~DeFanti, and L.~Smarr, ``Workflow-driven distributed
  machine learning in chase-ci: A cognitive hardware and software ecosystem
  community infrastructure,'' \emph{arXiv preprint arXiv:1903.06802}, 2019.

\bibitem{magri2014science}
D.~R.~C. Magri, T.~C.~M. de~Brito~Carvalho, F.~F. Redigolo, M.~A.~T. Rojas,
  M.~A.~S. Junior, L.~N. Ciuffo, G.~N. Dias, A.~S. de~Moura, and F.~Vetter,
  ``Science dmz: Support for e-science in brazil,'' in \emph{2014 IEEE 10th
  International Conference on e-Science}, vol.~2.\hskip 1em plus 0.5em minus
  0.4em\relax IEEE, 2014, pp. 75--78.

\bibitem{calyam2014wide}
P.~Calyam, A.~Berryman, E.~Saule, H.~Subramoni, P.~Schopis, G.~Springer,
  U.~Catalyurek, and D.~K. Panda, ``Wide-area overlay networking to manage
  science dmz accelerated flows,'' in \emph{2014 International Conference on
  Computing, Networking and Communications (ICNC)}.\hskip 1em plus 0.5em minus
  0.4em\relax IEEE, 2014, pp. 269--275.

\bibitem{farrell2016science}
\BIBentryALTinterwordspacing
L.~Farrell, ``{Science DMZ: The fast path for science data},'' \emph{Sci.
  Node}, May 2016. [Online]. Available:
  \url{https://sciencenode.org/feature/science-dmz-a-data-highway-system.php}
\BIBentrySTDinterwordspacing

\bibitem{qin2019towards}
Y.~Qin, A.~Simonet, P.~E. Davis, A.~Nouri, Z.~Wang, P.~Manish, and I.~Rodero,
  ``Towards a smart, internet-scale cache service for data intensive scientific
  applications,'' in \emph{Proceedings of the 10th Workshop on Scientific Cloud
  Computing}.\hskip 1em plus 0.5em minus 0.4em\relax ACM, 2019, pp. 11--18.

\bibitem{xiong2016prefetching}
L.~Xiong, Z.~Xu, H.~Wang, S.~Jia, and L.~Zhu, ``Prefetching scheme for massive
  spatiotemporal data in a smart city,'' \emph{International Journal of
  Distributed Sensor Networks}, vol.~12, no.~1, p. 4127358, 2016.

\bibitem{li2012prefetching}
R.~Li, R.~Guo, Z.~Xu, and W.~Feng, ``A prefetching model based on access
  popularity for geospatial data in a cluster-based caching system,''
  \emph{International Journal of Geographical Information Science}, vol.~26,
  no.~10, pp. 1831--1844, 2012.

\bibitem{ligo}
``{LIGO: Laser Interferometer Gravitational-Wave Observatory},''
  \url{https://www.ligo.caltech.edu/mit/}.

\bibitem{ska_web}
``{SKA: Square Kilometer Array},''
  \url{https://www.skatelescope.org/the-ska-project/}.

\bibitem{2019_LF}
``{2019 NSF Workshop on Connecting Large Facilities and Cyberinfrastructure},''
  \url{https://facilitiesci.github.io/2019/}.

\bibitem{dart2017assessment}
E.~Dart, M.~F. Wehner \emph{et~al.}, ``An assessment of data transfer
  performance for large-scale climate data analysis and recommendations for the
  data infrastructure for cmip6,'' \emph{arXiv preprint arXiv:1709.09575},
  2017.

\bibitem{2019_ci_blueprint}
``{Transforming Science Through Cyberinfrastructure},''
  \url{https://www.nsf.gov/cise/oac/vision/blueprint-2019/Overview-Computational.pdf},
  2019.

\bibitem{crichigno2018comprehensive}
J.~Crichigno, E.~Bou-Harb, and N.~Ghani, ``A comprehensive tutorial on science
  dmz,'' \emph{IEEE Communications Surveys \& Tutorials}, vol.~21, no.~2, pp.
  2041--2078, 2018.

\bibitem{dart2014science}
E.~Dart, L.~Rotman, B.~Tierney, M.~Hester, and J.~Zurawski, ``The science dmz:
  A network design pattern for data-intensive science,'' \emph{Scientific
  Programming}, vol.~22, no.~2, pp. 173--185, 2014.

\bibitem{chard2016globus}
K.~Chard, S.~Tuecke, and I.~Foster, ``Globus: Recent enhancements and future
  plans,'' in \emph{Proceedings of the XSEDE16 Conference on Diversity, Big
  Data, and Science at Scale}, 2016, pp. 1--8.

\bibitem{allcock2005globus}
W.~Allcock, J.~Bresnahan, R.~Kettimuthu, and M.~Link, ``The globus striped
  gridftp framework and server,'' in \emph{SC'05: Proceedings of the 2005
  ACM/IEEE conference on Supercomputing}.\hskip 1em plus 0.5em minus
  0.4em\relax IEEE, 2005, pp. 54--54.

\bibitem{radic2007optimization}
B.~Radi{\'c}, V.~Kaji{\'c}, and E.~Imamagi{\'c}, ``Optimization of data
  transfer for grid using gridftp,'' \emph{Journal of computing and information
  technology}, vol.~15, no.~4, pp. 347--353, 2007.

\bibitem{liu2018toward}
Z.~Liu, R.~Kettimuthu, I.~Foster, and P.~H. Beckman, ``Toward a smart data
  transfer node,'' \emph{Future Generation Computer Systems}, vol.~89, pp.
  10--18, 2018.

\bibitem{kettimuthu2018transferring}
R.~Kettimuthu, Z.~Liu, D.~Wheeler, I.~Foster, K.~Heitmann, and F.~Cappello,
  ``Transferring a petabyte in a day,'' \emph{Future Generation Computer
  Systems}, vol.~88, pp. 191--198, 2018.

\bibitem{kroeger1997exploring}
T.~M. Kroeger, D.~D. Long, J.~C. Mogul \emph{et~al.}, ``Exploring the bounds of
  web latency reduction from caching and prefetching.'' in \emph{USENIX
  Symposium on Internet Technologies and Systems}, 1997, pp. 13--22.

\bibitem{pallis2008clustering}
G.~Pallis, A.~Vakali, and J.~Pokorny, ``A clustering-based prefetching scheme
  on a web cache environment,'' \emph{Computers \& Electrical Engineering},
  vol.~34, no.~4, pp. 309--323, 2008.

\bibitem{huang2008mining}
Y.-F. Huang and J.-M. Hsu, ``Mining web logs to improve hit ratios of
  prefetching and caching,'' \emph{Knowledge-Based Systems}, vol.~21, no.~1,
  pp. 62--69, 2008.

\bibitem{mokhtarian2014caching}
K.~Mokhtarian and H.-A. Jacobsen, ``Caching in video cdns: Building strong
  lines of defense,'' in \emph{Proceedings of the ninth European conference on
  computer systems}, 2014, pp. 1--13.

\bibitem{podlipnig2003survey}
S.~Podlipnig and L.~B{\"o}sz{\"o}rmenyi, ``A survey of web cache replacement
  strategies,'' \emph{ACM Computing Surveys (CSUR)}, vol.~35, no.~4, pp.
  374--398, 2003.

\bibitem{vakali2000lru}
A.~Vakali, ``Lru-based algorithms for web cache replacement,'' in
  \emph{International conference on electronic commerce and web
  technologies}.\hskip 1em plus 0.5em minus 0.4em\relax Springer, 2000, pp.
  409--418.

\bibitem{cherkasova2001role}
L.~Cherkasova and G.~Ciardo, ``Role of aging, frequency, and size in web cache
  replacement policies,'' in \emph{International Conference on High-Performance
  Computing and Networking}.\hskip 1em plus 0.5em minus 0.4em\relax Springer,
  2001, pp. 114--123.

\bibitem{jin2001greedydual}
S.~Jin and A.~Bestavros, ``Greedydual web caching algorithm: exploiting the
  two sources of temporal locality in web request streams,'' \emph{Computer
  Communications}, vol.~24, no.~2, pp. 174--183, 2001.

\bibitem{cao1997cost}
P.~Cao and S.~Irani, ``Cost-aware www proxy caching algorithms.'' in
  \emph{Usenix symposium on internet technologies and systems}, vol.~12,
  no.~97, 1997, pp. 193--206.

\bibitem{ali2012intelligent}
W.~Ali, S.~M. Shamsuddin, and A.~S. Ismail, ``Intelligent web proxy caching
  approaches based on machine learning techniques,'' \emph{Decision Support
  Systems}, vol.~53, no.~3, pp. 565--579, 2012.

\bibitem{ali2014performance}
W.~Ali, S.~Sulaiman, and N.~Ahmad, ``Performance improvement of
  least-recently-used policy in web proxy cache replacement using supervised
  machine learning,'' \emph{International Journal of Advances in Soft Computing
  \& Its Applications}, vol.~6, no.~1, 2014.

\bibitem{xu2004keyword}
C.-Z. Xu and T.~I. Ibrahim, ``A keyword-based semantic prefetching approach in
  internet news services,'' \emph{IEEE Transactions on Knowledge and Data
  Engineering}, vol.~16, no.~5, pp. 601--611, 2004.

\bibitem{yang2001mining}
Q.~Yang, H.~H. Zhang, and T.~Li, ``Mining web logs for prediction models in www
  caching and prefetching,'' in \emph{Proceedings of the seventh ACM SIGKDD
  international conference on Knowledge discovery and data mining}, 2001, pp.
  473--478.

\bibitem{nanopoulos2003data}
A.~Nanopoulos, D.~Katsaros, and Y.~Manolopoulos, ``A data mining algorithm for
  generalized web prefetching,'' \emph{IEEE transactions on knowledge and data
  engineering}, vol.~15, no.~5, pp. 1155--1169, 2003.

\bibitem{wong2006web}
K.-Y. Wong, ``Web cache replacement policies: a pragmatic approach,''
  \emph{IEEE Network}, vol.~20, no.~1, pp. 28--34, 2006.

\bibitem{ooi-tos}
L.~M. Smith, J.~A. Barth, D.~S. Kelley, A.~Plueddemann, I.~Rodero, G.~A. Ulses,
  M.~F. Vardaro, and R.~Weller, ``The ocean observatories initiative,''
  \emph{Oceanography}, vol.~31, no.~1, pp. 16--35, 2018.

\bibitem{ooimartech16}
I.~Rodero and M.~Parashar, ``{Architecting the cyberinfrastructure for National
  Science Foundation Ocean Observatories Initiative (OOI)},'' \emph{{7th
  International Workshop on Marine Technology: MARTECH 2016}}, pp. 99--101,
  2016.

\bibitem{nsf_gage}
``{NSF Awards the Geodetic Facility for the Advancement of Geoscience (GAGE) to
  UNAVCO},'' \url{https://www.unavco.org/highlights/2018/award.html}.

\bibitem{ooi_2020}
``{OOI: Ocean Observatories Initiative},''
  \url{https://oceanobservatories.org/}.

\bibitem{contreras2003arima}
J.~Contreras, R.~Espinola, F.~J. Nogales, and A.~J. Conejo, ``Arima models to
  predict next-day electricity prices,'' \emph{IEEE transactions on power
  systems}, vol.~18, no.~3, pp. 1014--1020, 2003.

\bibitem{faruk2010hybrid}
D.~{\"O}. Faruk, ``A hybrid neural network and arima model for water quality
  time series prediction,'' \emph{Engineering applications of artificial
  intelligence}, vol.~23, no.~4, pp. 586--594, 2010.

\bibitem{han2000mining}
J.~Han, J.~Pei, and Y.~Yin, ``Mining frequent patterns without candidate
  generation,'' in \emph{ACM sigmod record}, vol.~29, no.~2.\hskip 1em plus
  0.5em minus 0.4em\relax ACM, 2000, pp. 1--12.

\bibitem{pan2017enhanced}
S.~Pan, Y.~Chong, Z.~Xu, and X.~Tan, ``An enhanced active caching strategy for
  data-intensive computations in distributed gis,'' \emph{The Journal of
  Supercomputing}, vol.~73, no.~10, pp. 4324--4346, 2017.

\bibitem{li2017replication}
R.~Li, W.~Feng, H.~Wu, and Q.~Huang, ``A replication strategy for a distributed
  high-speed caching system based on spatiotemporal access patterns of
  geospatial data,'' \emph{Computers, Environment and Urban Systems}, vol.~61,
  pp. 163--171, 2017.

\bibitem{xiong2018replication}
L.~Xiong, L.~Yang, Y.~Tao, J.~Xu, and L.~Zhao, ``Replication strategy for
  spatiotemporal data based on distributed caching system,'' \emph{Sensors},
  vol.~18, no.~1, p. 222, 2018.

\bibitem{tseng2019towards}
S.-M. Tseng, B.~Nicolae, G.~Bosilca, E.~Jeannot, A.~Chandramowlishwaran, and
  F.~Cappello, ``Towards portable online prediction of network utilization
  using mpi-level monitoring,'' in \emph{European Conference on Parallel
  Processing}.\hskip 1em plus 0.5em minus 0.4em\relax Springer, 2019, pp.
  47--60.

\end{thebibliography}


\end{document}